\documentclass{jpsj3}
\usepackage{type1cm}
\usepackage{tabularx}
\usepackage{lscape}
\usepackage{color}

\title{
Theoretical Study on Anisotropic Magnetoresistance Effects 
of Arbitrary Directions of Current and Magnetization for Ferromagnets: 
Application to Transverse Anisotropic Magnetoresistance Effect
}
\author{Satoshi Kokado$^1$\thanks{E-mail address: 
kokado.satoshi@shizuoka.ac.jp
} and Masakiyo Tsunoda$^{2,3}$ 
}
\inst{
$^1$Electronics and Materials Science Course, 
Department of Engineering, 
Graduate School of Integrated Science and Technology, 
Shizuoka University, Hamamatsu 432-8561, Japan \\
$^2$Department of Electronic Engineering, 
Graduate School of Engineering, Tohoku University, 
Sendai 980-8579, Japan\\
$^3$Center for Spintronics Research Network, Tohoku University, 
Sendai 980-8577, Japan
}
\abst{
We develop a theory of 
the anisotropic magnetoresistance (AMR) effects 
of arbitrary directions of current and magnetization for ferromagnets. 
Here, we use the electron scattering theory 
with 
the $s$--$s$ and $s$--$d$ scattering processes, 
where $s$ is the conduction electron state 
and $d$ is the localized d states. 
The resistivity due to electron scattering 
is expressed by 
the probability density of the d states of the current direction. 
The d states are numerically obtained by 
applying the exact diagonalization method 
to the Hamiltonian of the d states 
with the exchange field, crystal field, and spin--orbit interaction. 
Using the theory, 
we investigate the transverse AMR (TAMR) effect 
for strong ferromagnets 
with a crystal field of cubic or tetragonal symmetry. 
The cubic systems 
exhibit the fourfold symmetric TAMR effect, 
whereas 
the tetragonal systems 
show the twofold and fourfold symmetric TAMR effect. 
On the basis of the above results, 
we also comment on the experimental results of 
the TAMR effect for Fe$_4$N.

}

\begin{document}
\maketitle

\section{Introduction}
\begin{figure}[ht]
\begin{center}
\includegraphics[width=0.6\linewidth]{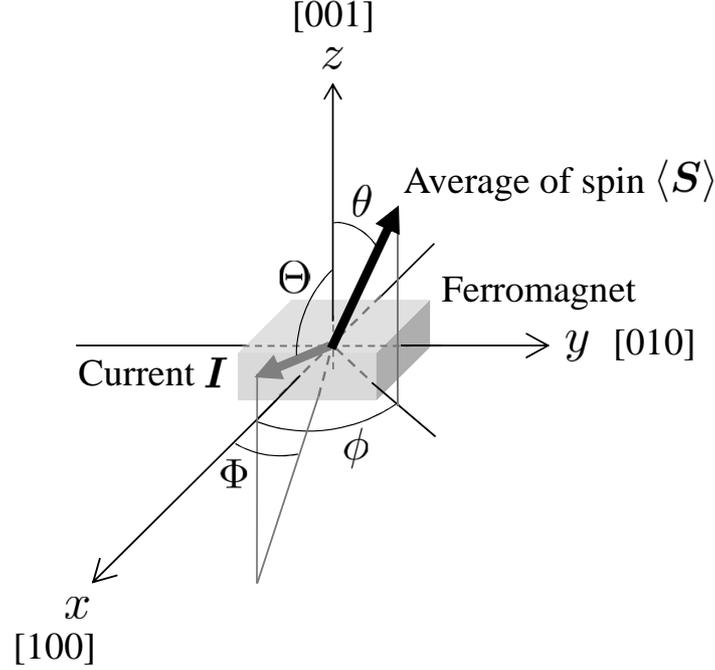}
\caption{
Sketch of sample geometry 
for AMR measurement. 
The current ${\mbox{\boldmath $I$}}$ flows 
in the direction indexed by $\Theta$ and $\Phi$, 
%where $\Theta$ and $\Phi$ 
which 
are the polar angle and azimuthal angle 
of ${\mbox{\boldmath $I$}}$, respectively. 
The thermal average of the spin 
$\langle {\mbox{\boldmath $S$}} \rangle$ 
($\propto$$-{\mbox{\boldmath $M$}}$) 
is oriented in the direction indexed by $\theta$ and $\phi$, 
%where 
%$\theta$ and $\phi$ 
which are 
the polar angle and azimuthal angle 
of 
$\langle {\mbox{\boldmath $S$}} \rangle$, 
respectively. 
In addition, the $x$-, $y$-, and $z$-axes are specified 
to describe the Hamiltonian of Eq. (\ref{Hamiltonian}). 
}
\label{sample}
\end{center}
\end{figure}

The anisotropic magnetoresistance (AMR) effect for ferromagnets 
is a fundamental phenomenon in which the electrical resistivity depends on 
the direction of magnetization ${\mbox{\boldmath $M$}}$.\cite{Thomson,McGuire,Campbell,Berger,Miyazaki,Tsunoda1,Tsunoda2,Tsunoda3,Kabara2,Yang,Sakuraba,Kokado1,Kokado2,Kokado3,Kokado4,Kokado5,Kokado6,Zhao,Sato1,Sato2} 
Figure \ref{sample} shows the sample geometry 
for the AMR measurement. 
The current ${\mbox{\boldmath $I$}}$ flows 
in the direction indexed by $\Theta$ and $\Phi$, 
where 
$\Theta$ and $\Phi$ 
are the polar angle and azimuthal angle 
of ${\mbox{\boldmath $I$}}$, respectively. 
The thermal average of the spin 
$\langle {\mbox{\boldmath $S$}} \rangle$ 
($\propto$$-{\mbox{\boldmath $M$}}$) 
is oriented 
in the direction indexed by $\theta$ and $\phi$, 
where 
$\theta$ and $\phi$ are 
the polar angle and azimuthal angle 
of 
$\langle {\mbox{\boldmath $S$}} \rangle$, 
respectively. 
The efficiency of the effect, the ``AMR ratio'', 
is defined by
\begin{eqnarray}
\label{amr^i(theta,phi)}
{\rm AMR}(\Theta,\Phi;\theta,\phi)
= \frac{\rho (\Theta,\Phi;\theta,\phi) - \rho (\Theta,\Phi;\theta_0,\phi_0)}{\rho (\Theta,\Phi;\theta_0,\phi_0)}. 
\end{eqnarray}
Here, 
$\rho (\Theta,\Phi;\theta,\phi)$ represents resistivity at 
$(\Theta, \Phi)$ 
and 
$(\theta, \phi)$, 
as will be given by Eq. (\ref{rho^im}). 
In addition, 
$\theta_0$ and $\phi_0$ are 
the specific $\theta$ and $\phi$, which are chosen as the reference direction.

Usually, 
the AMR ratio is 
%has been 
experimentally measured 
for the in-plane configuration, 
in which 
${\mbox{\boldmath $I$}}$ flows 
in the [100] or [110] direction 
and %the magnetization 
${\mbox{\boldmath $M$}}$ lies in the (001) plane.\cite{Tsunoda1,Tsunoda2,Sato1,Sato2} 
In addition, the in-plane AMR ratios have often been analyzed 
by using analytic expressions of the AMR ratio 
derived from the electron scattering theory.\cite{Kokado1,Kokado2,Kokado3,Kokado4,Kokado5,Kokado6} 
This theory has 
taken into account 
all the electron scattering processes 
from the conduction state to the localized d states 
through a nonmagnetic impurity. 
The wave functions of the d states have been analytically obtained 
within the perturbation theory.

%Nowadays, 
Currently, AMR effects are being investigated for various directions of 
${\mbox{\boldmath $I$}}$ and ${\mbox{\boldmath $M$}}$. 
For example, Kabara {\it et al.}\cite{Kabara1,Tsunoda4} 
experimentally observed the transverse AMR (TAMR) effect 
for Fe$_4$N, 
where Fe$_4$N is considered to be 
a strong ferromagnet.\cite{SW_FM,Kokado1,Kokado3} 
In this effect, 
${\mbox{\boldmath $I$}}$ flows in the [100] direction 
(i.e., $\Theta=\pi/2$ and $\Phi=0$) 
and ${\mbox{\boldmath $M$}}$ lies in the (100) plane 
(i.e., an arbitrary $\theta$ and $\phi= \pi/2$). 
The TAMR ratio is defined by
\begin{eqnarray}
\label{TAMR}
{\rm TAMR}(\theta) = 
\frac{\rho (\theta) - \rho (0)}
{\rho (0)}, 
\end{eqnarray}
with 
$\rho(\theta) = \rho (\pi/2,0; \theta,\pi/2)$, 
where 
$\theta=0$ is chosen as the reference direction. 
This TAMR$(\theta)$ of Eq. (\ref{TAMR}) 
is generally expressed as
\begin{eqnarray}
\label{Tr-AMR(theta)}
{\rm TAMR}(\theta) = C_0 + C_2 \cos 2\theta + C_4 \cos 4\theta,
\end{eqnarray}
where $C_0$ is a constant term and 
$C_2$ ($C_4$) is the coefficient of the twofold (fourfold) symmetric term. 
They reported that 
Fe$_4$N exhibits 
the enhancement of $|C_2|$ and $|C_4|$ 
for $T \lesssim 35$ K, 
i.e., 
a change from 
the fourfold symmetric TAMR($\theta$) 
with 
$C_2=0$ and $C_4=-0.005$ at $T \sim 35$ K 
to 
the twofold and fourfold symmetric TAMR($\theta$) 
with 
$C_2=-0.02$ and $C_4=-0.01$ at $T=5$ K.\cite{Kabara1,Tsunoda4} 
In addition, 
using the multi-orbital d-impurity Anderson model, 
Yahagi {\it et al.}\cite{Yahagi} 
confirmed 
that the system with the crystal field of cubic symmetry 
shows 
the fourfold symmetric TAMR($\theta$) 
with $C_2 =0$ and $C_4 \ne 0$. 
Such a TAMR effect has also been 
measured to separate the contributions from 
the TAMR effect 
and the spin Hall magnetoresistance effect 
in 
the whole magnetoresistance effect 
for a nonmagnet/ferrimagnet structure,\cite{Zhou} 
for example. 
%and so on. 

In the future, 
AMR effects will be examined more extensively for various directions of 
${\mbox{\boldmath $I$}}$ and ${\mbox{\boldmath $M$}}$. 
On the other hand, 
theories to systematically investigate various AMR effects 
have seldom been proposed. 
In particular, 
the twofold and fourfold TAMR effect 
has rarely been 
investigated theoretically. 
We also note that 
it may be difficult to obtain 
an analytic expression of 
the AMR ratio 
of arbitrary directions of 
${\mbox{\boldmath $I$}}$ and ${\mbox{\boldmath $M$}}$ 
or TAMR$(\theta)$ 
by the perturbation theory.

In this paper, 
we first develop a theory of 
the AMR effects of arbitrary directions of 
${\mbox{\boldmath $I$}}$ and ${\mbox{\boldmath $M$}}$ for ferromagnets 
by taking 
into account 
all the electron scattering processes 
from the conduction state to the localized d states. 
The wave functions of the d states are numerically obtained 
by using the exact diagonalization method for the Hamiltonian 
of the d states with 
the exchange field, crystal field, and spin--orbit interaction. 
Using this theory, 
we next investigate the TAMR effect for strong ferromagnets. 
We find that 
the system with the crystal field of cubic symmetry exhibits 
the fourfold symmetric TAMR effect, 
while 
the system with the crystal field of tetragonal symmetry 
shows 
the twofold and fourfold symmetric TAMR effect. 
Finally, on the basis of the above results, 
we comment on the experimental results of 
the TAMR effect for Fe$_4$N.

The present paper is organized as follows. 
In Sect. \ref{sec_theory}, 
we describe a theory 
of the AMR effects of arbitrary directions of 
${\mbox{\boldmath $I$}}$ and ${\mbox{\boldmath $M$}}$ 
for ferromagnets. 
In Sect. \ref{sec_appl}, 
we apply the theory 
to the TAMR effects. 
In Sect. \ref{strong}, 
we investigate 
the TAMR effects for strong ferromagnets 
with a crystal field of cubic or tetragonal symmetry. 
In Sect. \ref{comment}, 
we comment on the experimental results of 
the TAMR effect for Fe$_4$N. 
The conclusion is presented in Sect. \ref{sec_conc}. 
In Appendix \ref{relation}, 
we show the relation between 
TAMR$(\theta)$ 
and 
the probability density of the d states 
of the ${\mbox{\boldmath $I$}}$ direction 
for strong ferromagnets. 
Appendix \ref{TAMR_egeg} 
gives the approximate expression
of TAMR($\theta$) 
for strong ferromagnets. 
In Appendix \ref{cond1}, 
we approximately obtain 
the condition for 
the probability density of the d states 
of the ${\mbox{\boldmath $I$}}$ direction.

\section{Theory}
\label{sec_theory}
In this section, we describe the electron scattering theory 
to obtain 
$\rho(\Theta,\Phi;\theta,\phi)$ 
for ferromagnets. 
This theory is developed on the basis of our previous study.
\cite{Kokado1,Kokado2,Kokado3,Kokado4,Kokado5,Kokado6} 
Here, we use the two-current model 
with the $s$--$s$ and $s$--$d$ scatterings 
for the system shown in Fig. \ref{sample}, 
where 
$s$ is the conduction electron state 
and $d$ is the localized d states.
\cite{Kokado1,Kokado2,Kokado3,Kokado4,Kokado5,Kokado6}

\subsection{Hamiltonian}
\label{sub_Ham}
To obtain the localized d states, 
we give the Hamiltonian ${\cal H}$ of the localized d states 
of a single atom in a ferromagnet with 
orbital energies, 
an exchange field, 
and a spin--orbit interaction. 
The Hamiltonian ${\cal H}$ is expressed as
\begin{eqnarray}
\label{Hamiltonian}
&&{\cal H} = {\cal H}_{\rm orbital} 
+ {\cal H}_{\rm ex} 
+ {\cal H}_{\rm so}, 
\end{eqnarray}
where
\begin{eqnarray}
&&
{\cal H}_{\rm orbital}
=
\sum_{\sigma=+,-}
\Bigg( 
E_{xy} 
|xy, \chi_\sigma (\theta,\phi) \rangle \langle xy, \chi_\sigma (\theta,\phi)| 
+ 
E_{yz} 
|yz, \chi_\sigma (\theta,\phi) \rangle \langle yz, \chi_\sigma (\theta,\phi)| \nonumber \\
&&\hspace*{1.8cm}+ E_{xz} 
|xz, \chi_\sigma (\theta,\phi) \rangle 
\langle xz, \chi_\sigma (\theta,\phi)|  \nonumber \\
&& \hspace*{1.8cm}+ 
E_{x^2-y^2}
|x^2-y^2, \chi_\sigma (\theta,\phi)\rangle \langle x^2-y^2, \chi_\sigma (\theta,\phi)| \nonumber \\
&& \hspace*{1.8cm}+ 
E_{3z^2-r^2}
|3z^2-r^2, \chi_\sigma (\theta,\phi) \rangle \langle 3z^2-r^2, \chi_\sigma (\theta,\phi)| 
\Bigg), \\
&&{\cal H}_{\rm ex} = - {\mbox{\boldmath $S$}} \cdot {\mbox{\boldmath $H$}}, \\
&&{\cal H}_{\rm so} = \lambda (L_x S_x + L_y S_y) + \lambda' L_z S_z,
%{\mbox{\boldmath $L$}} \cdot {\mbox{\boldmath $S$}}, 
\end{eqnarray}
with
\begin{eqnarray}
&&{\mbox{\boldmath $S$}}=(S_x, S_y, S_z), \\
&&{\mbox{\boldmath $L$}}=(L_x, L_y, L_z), \\
&&{\mbox{\boldmath $H$}}=
H \left(\sin \theta \cos \phi, \sin \theta \sin \phi, \cos \theta \right). 
\end{eqnarray}
The above terms are explained as follows. 
The term 
${\cal H}_{\rm ex}$ 
is the 
%Zeeman 
interaction between 
the spin angular momentum ${\mbox{\boldmath $S$}}$ 
and 
the exchange field of the ferromagnet 
${\mbox{\boldmath $H$}}$, 
where 
${\mbox{\boldmath $H$}} \propto -{\mbox{\boldmath $M$}}$, 
${\mbox{\boldmath $H$}} \propto \langle {\mbox{\boldmath $S$}} \rangle$, 
and $H > 0$. 
The term 
${\cal H}_{\rm so}$ is the spin--orbit interaction, 
where 
${\mbox{\boldmath $L$}}$ is the orbital angular momentum, 
$\lambda$ is the spin--orbit coupling constant 
of the $L_x S_x + L_y S_y$ operator, 
%(i.e., the mixing ), 
and 
$\lambda'$ is the spin--orbit coupling constant 
of the $L_z S_z$ operator. 
%where 
The different symbols $\lambda$ and $\lambda'$ are introduced to 
%The constants $\lambda$ and $\lambda'$ are introduced to 
%find 
determine the contributions of the respective terms to the d states, 
where they are unique in a spherically symmetric potential. 
%the resistivity. 
%the AMR effect. 
%and 
%${\mbox{\boldmath $L$}}$ is the orbital angular momentum. 
The spin quantum number $S$ 
and the azimuthal quantum number $L$ 
are chosen to be 
$S=1/2$ and $L=2$.\cite{Kokado1} 
The term 
${\cal H}_{\rm orbital}$ represents the orbital energies. 
The quantity 
$E_i$ is the energy level of the $i$ orbital, 
where 
$i = xy$, $yz$, $xz$, $x^2-y^2$, and $3z^2-r^2$. 
The state $|i,\chi_\sigma (\theta,\phi) \rangle$ is expressed by 
$|i,\chi_\sigma (\theta,\phi) \rangle=|i \rangle |\chi_\sigma (\theta,\phi) \rangle$. 
The state $|i \rangle$ is the orbital state, 
defined by 
$|xy \rangle = xyf(r)$, 
$|yz \rangle = yzf(r)$, 
$|xz \rangle = xzf(r)$, 
$|x^2-y^2 \rangle = (1/2)(x^2-y^2)f(r)$, and 
$|3z^3-r^2 \rangle = [1/(2\sqrt{3})](3z^2-r^2)f(r)$, 
with $r=\sqrt{x^2 + y^2 + z^2}$ 
and $f(r) = \Gamma e^{-\zeta r}$, 
where 
$f(r)$ is the radial part of the d orbital, and 
$\Gamma$ and $\zeta$ are constants. 
The state 
$|\chi_\sigma (\theta,\phi) \rangle$ ($\sigma=+$ and $-$) is the spin state, i.e., 
\begin{eqnarray}
\label{+spin}
&&|\chi_+ (\theta,\phi) \rangle =  e^{-i\phi} \cos \frac{\theta}{2} |\uparrow \rangle + \sin \frac{\theta}{2} |\downarrow \rangle, \\
\label{-spin}
&&|\chi_- (\theta,\phi) \rangle = -e^{-i\phi} \sin \frac{\theta}{2} |\uparrow \rangle + \cos \frac{\theta}{2}|\downarrow \rangle, 
\end{eqnarray}
which are eigenstates 
of ${\cal H}_{\rm ex}$. 
Here, $|\chi_+ (\theta,\phi) \rangle$ ($|\chi_- (\theta,\phi) \rangle$) 
denotes the up spin state (down spin state) 
for the case in which 
the quantization axis is chosen 
along the $\langle {\mbox{\boldmath $S$}} \rangle$ direction. 
The state 
$|\uparrow \rangle$ ($|\downarrow \rangle$) represents 
the up spin state (down spin state) 
for the case in which 
the quantization axis is chosen along 
the $z$-axis. 
In addition, 
this system is 
a normalized orthogonal system 
with 
$\langle i' | i \rangle = \delta_{i',i}$ 
and 
$\langle \chi_{\sigma'} (\theta,\phi) | \chi_\sigma (\theta,\phi) \rangle = \delta_{\sigma',\sigma}$. 
Table \ref{matrix} shows the matrix element of ${\cal H}$ 
expressed by 
the basis set of 
$|xy, \chi_\pm (\theta,\phi) \rangle$, 
$|yz, \chi_\pm (\theta,\phi) \rangle$, 
$|xz, \chi_\pm (\theta,\phi) \rangle$, 
$|x^2 - y^2, \chi_\pm (\theta,\phi) \rangle$, 
and $|3z^2 -r^2, \chi_\pm (\theta,\phi) \rangle$.

\begin{landscape}
\begin{table*}[ht]
\caption{Matrix element of ${\cal H}$ of Eq. (\ref{Hamiltonian}) 
expressed by 
the basis set of 
$|xy, \chi_\pm (\theta,\phi) \rangle$, 
$|yz, \chi_\pm (\theta,\phi) \rangle$, 
$|xz, \chi_\pm (\theta,\phi) \rangle$, 
$|x^2 - y^2, \chi_\pm (\theta,\phi) \rangle$, 
and $|3z^2 -r^2, \chi_\pm (\theta,\phi) \rangle$. 
The system with the crystal field of tetragonal symmetry 
in Fig. \ref{energy} 
has 
$E_{xy}=0$, 
$E_{xz}=E_{yz}=\delta_\varsigma$, 
$E_{x^2-y^2}=\Delta$, 
and 
$E_{3z^2-r^2}=\Delta+\delta_\gamma$. 
The system with the crystal field of cubic symmetry 
has 
$E_{xy}=E_{xz}=E_{yz}=0$ 
and 
$E_{x^2-y^2}=E_{3z^2-r^2}=\Delta$. 
In addition, 
%In 
the case of the TAMR effect 
%We also set 
has $\phi=\pi/2$. 
%in investigating the TAMR effect. 
}
\begin{center}
\scalebox{0.65}{
\begin{tabular}{|c||c|c|c|c|c|c|c|c|c|c|}
\hline 
   & $|xy,\chi_+ \rangle$ & $|yz,\chi_+ \rangle$ & $|xz,\chi_+ \rangle$ & $|xy,\chi_- \rangle$ & $|yz,\chi_- \rangle$ & $|xz,\chi_- \rangle$ & $|x^2-y^2, \chi_+ \rangle$ & $|3z^2-r^2, \chi_+ \rangle$ & $|x^2-y^2, \chi_- \rangle$ & $|3z^2-r^2, \chi_- \rangle$ \\
\hline 
\hline 
$\langle xy,\chi_+ |$ & 
$E_{xy}- \frac{H}{2}$ & $i\frac{\lambda}{2}\sin \theta \sin \phi$ & $-i\frac{\lambda}{2}\sin \theta\cos \phi$ & 0 & $\frac{\lambda}{2}(e^{-i\phi} \sin^2 \frac{\theta}{2} $ & $i\frac{\lambda}{2} (e^{-i\phi} \sin^2 \frac{\theta}{2} $ & $i\lambda' \cos \theta$ & 0 & $- i \lambda' \sin \theta$ & 0 \\
& &  &  &  & $+ e^{i\phi} \cos^2 \frac{\theta}{2})$ & $ - e^{i\phi} \cos^2 \frac{\theta}{2})$ &  &  &  &  \\
\hline 
$\langle yz,\chi_+ |$ & 
$-i\frac{\lambda}{2} \sin \theta \sin \phi$ & $E_{yz} - \frac{H}{2}$ & $i\frac{\lambda'}{2} \cos \theta$ & $- \frac{\lambda}{2} (e^{-i\phi} \sin^2 \frac{\theta}{2} $ & 0 & $-i \frac{\lambda'}{2}\sin \theta$ & $-i \frac{\lambda}{2} \sin \theta \cos \phi$ & $-i\frac{\sqrt{3}\lambda}{2} \sin \theta \cos \phi$ & $-i\frac{\lambda}{2}(-e^{-i\phi} \sin^2 \frac{\theta}{2} $ & $-i\frac{\sqrt{3}\lambda}{2}(-e^{-i\phi} \sin^2 \frac{\theta}{2} $ \\
 &  &  &  & $+ e^{i\phi} \cos^2 \frac{\theta}{2})$ &  &  &  &  & $+ e^{i\phi} \cos^2 \frac{\theta}{2})$ & $+ e^{i\phi} \cos^2 \frac{\theta}{2})$ \\
\hline 
$\langle xz,\chi_+ |$ & 
$i\frac{\lambda}{2} \sin \theta \cos \phi$ & $-i \frac{\lambda'}{2} \cos \theta$ & $E_{xz}- \frac{H}{2}$ & $-i\frac{\lambda}{2} (e^{-i\phi} \sin^2 \frac{\theta}{2} $ & $i \frac{\lambda'}{2} \sin \theta$ & 0 & $-i\frac{\lambda}{2} \sin \theta \sin \phi$ & $i \frac{\sqrt{3} \lambda}{2}\sin \theta \sin \phi$ & $-\frac{\lambda}{2} (e^{-i\phi} \sin^2 \frac{\theta}{2} $ & $\frac{\sqrt{3}\lambda}{2}(e^{-i\phi} \sin^2 \frac{\theta}{2}$ \\
 & &  &  & $ - e^{i\phi} \cos^2 \frac{\theta}{2})$ &  &  &  &  & $ + e^{i\phi} \cos^2 \frac{\theta}{2})$ & $+ e^{i\phi} \cos^2 \frac{\theta}{2})$ \\
\hline
$\langle xy,\chi_- |$ & 
0 & $-\frac{\lambda}{2}(e^{-i\phi} \cos^2 \frac{\theta}{2} $ & $-i\frac{\lambda}{2} (e^{-i\phi} \cos^2 \frac{\theta}{2} $ & $E_{xy}+\frac{H}{2}$ & $-i\frac{\lambda}{2} \sin \theta \sin \phi$ & $i\frac{\lambda}{2} \sin \theta \cos \phi$ & $-i \lambda' \sin \theta$ & 0 & $-i\lambda' \cos \theta$ & 0 \\
 & & $+ e^{i\phi} \sin^2 \frac{\theta}{2})$ & $- e^{i\phi} \sin^2 \frac{\theta}{2})$ &  &  &   &  &  & & \\
\hline
$\langle yz,\chi_- |$ & 
$\frac{\lambda}{2}(e^{-i\phi} \cos^2 \frac{\theta}{2} $ & 0 & $-i \frac{\lambda'}{2}\sin \theta$ & $i\frac{\lambda}{2}\sin \theta \sin \phi$ & $E_{yz} + \frac{H}{2}$ & $-i \frac{\lambda'}{2}\cos \theta$ & $-i\frac{\lambda}{2}(e^{-i\phi} \cos^2 \frac{\theta}{2}$ & $-i\frac{\sqrt{3}\lambda}{2}(e^{-i\phi} \cos^2 \frac{\theta}{2} $ & $i\frac{\lambda }{2} \sin \theta \cos \phi$ & $i\frac{\sqrt{3}\lambda }{2} \sin \theta \cos \phi$ \\
 & $+ e^{i\phi} \sin^2 \frac{\theta}{2})$ &  &  & &  &  & $- e^{i\phi} \sin^2 \frac{\theta}{2})$ & $- e^{i\phi} \sin^2 \frac{\theta}{2})$ &  &  \\
\hline
$\langle xz,\chi_- |$ & 
$i\frac{\lambda}{2} (e^{-i\phi} \cos^2 \frac{\theta}{2} $ & $i\frac{\lambda'}{2}\sin \theta$ & 0 & $-i\frac{\lambda}{2}\sin \theta \cos \phi$ & $i\frac{\lambda'}{2}\cos \theta$ & $E_{xz} + \frac{H}{2}$ & $\frac{\lambda}{2}(e^{-i\phi} \cos^2 \frac{\theta}{2}$ & $-\frac{\sqrt{3}\lambda}{2} (e^{-i\phi} \cos^2 \frac{\theta}{2} $ & $i \frac{\lambda}{2} \sin \theta \sin \phi$ & $-i\frac{\sqrt{3}\lambda}{2} \sin \theta \sin \phi$ \\
 & $- e^{i\phi} \sin^2 \frac{\theta}{2})$ &  &  & &  &  & $+ e^{i\phi} \sin^2 \frac{\theta}{2})$ & $ + e^{i\phi} \sin^2 \frac{\theta}{2})$ & & \\
\hline
$\langle x^2-y^2,\chi_+ |$ & 
$-i\lambda' \cos \theta$ & $i\frac{\lambda}{2}\sin \theta \cos \phi$ & $i\frac{\lambda}{2} \sin \theta \sin \phi$ & $i \lambda' \sin \theta$  & $i\frac{\lambda}{2} (-e^{-i\phi} \sin^2 \frac{\theta}{2} $ & $\frac{\lambda}{2} (e^{-i\phi} \sin^2 \frac{\theta}{2} $ & $E_{x^2-y^2}- \frac{H}{2}$ & 0 & 0 & 0 \\
 & &  &  &   & $ + e^{i\phi} \cos^2 \frac{\theta}{2})$ & $ + e^{i\phi} \cos^2 \frac{\theta}{2})$ &  &  &  &  \\
\hline
$\langle 3z^2-r^2,\chi_+ |$ & 
0 & $i\frac{\sqrt{3}\lambda}{2} \sin \theta \cos \phi$ & $-i\frac{\sqrt{3}\lambda}{2} \sin \theta \sin \phi$ & 0 & $i\frac{\sqrt{3}\lambda}{2}(-e^{-i\phi} \sin^2 \frac{\theta}{2} $ & $-\frac{\sqrt{3}\lambda}{2}(e^{-i\phi} \sin^2 \frac{\theta}{2} $ & 0 & $E_{3z^2-r^2}- \frac{H}{2} $ & 0 & 0 \\
 & &  &  &  & $ + e^{i\phi} \cos^2 \frac{\theta}{2})$ & $+ e^{i\phi} \cos^2 \frac{\theta}{2})$ &  &  &  &  \\
\hline
$\langle x^2-y^2,\chi_- |$ & 
$i \lambda' \sin \theta$ & $i\frac{\lambda }{2}(e^{-i\phi} \cos^2 \frac{\theta}{2} $ & $-\frac{\lambda}{2} (e^{-i\phi} \cos^2 \frac{\theta}{2} $ & $i\lambda' \cos \theta$ & $-i \frac{\lambda }{2} \sin \theta \cos \phi$ & $- i \frac{\lambda}{2} \sin \theta \sin \phi$ & 0 & 0 & $E_{x^2-y^2}+ \frac{H}{2}$ & 0  \\
 & & $- e^{i\phi} \sin^2 \frac{\theta}{2})$ & $+ e^{i\phi} \sin^2 \frac{\theta}{2})$ &  &  &  &  &  &   &   \\
\hline
$\langle 3z^2-r^2,\chi_- |$ & 
0 & $i\frac{\sqrt{3}\lambda }{2} (e^{-i\phi} \cos^2 \frac{\theta}{2} $ & $\frac{\sqrt{3}\lambda }{2} (e^{-i\phi} \cos^2 \frac{\theta}{2} $ & 0 & $-i\frac{\sqrt{3}\lambda }{2}\sin \theta \cos \phi$ & $i\frac{\sqrt{3}\lambda }{2} \sin \theta \sin \phi$ & 0 & 0 & 0 & $E_{3z^2-r^2}+ \frac{H}{2} $ \\
 & & $- e^{i\phi} \sin^2 \frac{\theta}{2})$ & $ + e^{i\phi} \sin^2 \frac{\theta}{2})$ &  &  &  &  &  &  &  \\
\hline
\end{tabular}
}
\end{center}
\label{matrix}
\end{table*} 
\end{landscape}

\subsection{Localized d states}
\label{d_states}
Applying the exact diagonalization method 
to ${\cal H}$ of Eq. (\ref{Hamiltonian}) (i.e., Table \ref{matrix}), 
we numerically obtain 
the wave function of the localized d states, 
$|\psi_{j,\varsigma} (\theta,\phi))$. 
%\textcolor{red}
%{
In this study we mainly consider the case in which 
%the spin--orbit interaction is smaller than the crystal field energy formed by $E_i$, 
$\lambda$ is smaller than $H$ and 
the crystal field energy, 
% formed by $E_i$, 
where 
the crystal field energy is 
%, for example, 
$E_{x^2 - y^2} - E_{xy}$ ($= \Delta$) 
as noted in Sect. \ref{model_param} (see also Fig. \ref{energy}). 
%$\Delta$ ($=E_{x^2 - y^2} - E_{xy}$) in Fig. \ref{sample}. 
%the energy level of $|x^2-y^2, \chi_\sigma (\theta)  \rangle$ measured from that of $|xy, \chi_\sigma (\theta)  \rangle$, $\Delta$, in Fig. \ref{sample}. 
%[e.g., the energy level of $|x^2-y^2, \chi_\sigma (\theta)  \rangle$ measured from that of $|xy, \chi_\sigma (\theta)  \rangle$, $\Delta$, in Fig. \ref{sample}]. 
In this case, 
%This 
$|\psi_{j,\varsigma} (\theta,\phi))$ is 
expressed as 
%}
\begin{eqnarray}
\label{|m,chi_s)}
&&|\psi_{j,\varsigma} (\theta,\phi))=
\sum_{\substack{i=xy, yz, xz\\x^2-y^2, 3z^2-r^2}} \sum_{\sigma=+, -}
c_{i,\sigma}^{j,\varsigma} (\theta,\phi) 
|i,\chi_\sigma (\theta,\phi) \rangle, 
\end{eqnarray}
with 
$j = xy$, $yz$, $xz$, $x^2-y^2$, and $3z^2-r^2$ 
and 
$\varsigma = +$ and $-$. 
The quantity 
$c_{i,\sigma}^{j,\varsigma} (\theta,\phi)$ is the probability amplitude of 
$|i,\chi_\sigma (\theta,\phi) \rangle$. 
Here, $j$ and $\varsigma$ respectively specify 
the orbital and spin indexes of the dominant state in 
$\sum_i \sum_\sigma c_{i,\sigma}^{j,\varsigma} (\theta,\phi) 
|i,\chi_\sigma (\theta,\phi) \rangle$. 
%\textcolor{red}
%{
We also mention the case in which 
%the spin--orbit interaction is 
$\lambda$ is smaller than $H$ 
but larger than the crystal field energy 
(e.g., the case of no crystal field 
in Sects. \ref{model_param} and \ref{no_crystal}). 
In this case, 
$j$ does not correspond well to 
$xy$, $yz$, $xz$, $x^2-y^2$, and $3z^2-r^2$. 
Therefore, we may assign a number to $j$ 
in descending order of the eigenvalue of ${\cal H}$, i.e., 
$j=1$, 2, 3, 4, and 5. 
%$j=1$, 2, 3, $\cdots$, 10.
%}

The state $|\psi_{j,\varsigma} (\theta,\phi))$ has 
a normalized orthogonal system: 
\begin{eqnarray}
\label{orthonormal}
(\psi_{j',\varsigma'} (\theta,\phi)
|\psi_{j,\varsigma} (\theta,\phi) )
= \delta_{j',j} \delta_{\varsigma',\varsigma}, 
\end{eqnarray}
where
\begin{eqnarray}
&&
(\psi_{j,\varsigma} (\theta,\phi)|\psi_{j,\varsigma} (\theta,\phi))
=
\sum_i \sum_\sigma |c_{i,\sigma}^{j,\varsigma} (\theta,\phi)|^2=1. 
\end{eqnarray}
Furthermore, we have the following condition: 
\begin{eqnarray}
\label{condition}
&&\hspace*{-1cm}\sum_{j} \sum_{\varsigma}
\left[ c_{i,\sigma}^{j,\varsigma} (\theta,\phi) \right]^*
c_{i',\sigma'}^{j,\varsigma} (\theta,\phi) \nonumber \\
&&=\sum_{j} \sum_{\varsigma}
\langle i,\chi_\sigma (\theta,\phi) |\psi_{j,\varsigma} (\theta,\phi) )^*
\langle i',\chi_{\sigma'} (\theta,\phi) 
|\psi_{j,\varsigma} (\theta,\phi) ) \nonumber \\
&&=\sum_{j} \sum_{\varsigma}
\langle i',\chi_{\sigma'} (\theta,\phi) |\psi_{j,\varsigma} (\theta,\phi) )
(\psi_{j,\varsigma} (\theta,\phi) | i,\chi_\sigma (\theta,\phi) \rangle \nonumber \\
&&=\langle i',\chi_{\sigma'} (\theta,\phi) | i,\chi_\sigma (\theta,\phi) \rangle\nonumber \\
&&= \delta_{i',i}\delta_{\sigma',\sigma},
\end{eqnarray}
where 
the completeness, 
$\sum_j \sum_{\varsigma} |\psi_{j,\varsigma} (\theta,\phi) )(\psi_{j,\varsigma} (\theta,\phi) |=1$, 
has been used. 
In particular, 
when $i'=i$ and $\sigma'=\sigma$, 
Eq. (\ref{condition}) becomes 
\begin{eqnarray}
\label{111}
&&\sum_j \sum_\varsigma |c_{i,\sigma}^{j,\varsigma} (\theta,\phi)|^2=1.
\end{eqnarray}
Equations (\ref{condition}) and (\ref{111}) 
will be used in Eqs. (\ref{p_const}) and (\ref{p_const1}). 
%, respectively. 
%will be used in Eqs. (\ref{p_const}) and (\ref{222}), respectively. 

\subsection{Resistivity}
\label{sec_resistivity}
Using $|\psi_{j,\varsigma} (\theta,\phi))$ of Eq. (\ref{|m,chi_s)}), 
we can obtain an expression of $\rho (\Theta,\Phi;\theta,\phi)$. 
The resistivity 
$\rho (\Theta,\Phi;\theta,\phi)$ is described 
by the two-current model,\cite{Campbell} i.e.,
\begin{eqnarray}
\label{rho^im}
&&\rho (\Theta,\Phi;\theta,\phi) 
= \frac{ \rho_+ (\Theta,\Phi;\theta,\phi) \rho_- (\Theta,\Phi;\theta,\phi)}
{\rho_+ (\Theta,\Phi;\theta,\phi) + \rho_- (\Theta,\Phi;\theta,\phi)},
\end{eqnarray}
where 
$\rho_\sigma (\Theta,\Phi;\theta,\phi)$ 
is the resistivity of the $\sigma$ spin at 
($\Theta,\Phi$) and ($\theta,\phi$). 
Here, 
$\sigma=+$ ($-$) denotes the up spin (down spin) 
for the case in which 
the quantization axis is chosen along the direction of 
$\langle {\mbox{\boldmath $S$}} \rangle$ 
[see Eqs. (\ref{+spin}) and (\ref{-spin})]. 
The resistivity $\rho_\sigma (\Theta,\Phi;\theta,\phi)$ is written as 
\begin{eqnarray}
\label{rho_sigma^i}
&&\rho_\sigma (\Theta,\Phi;\theta,\phi)=
\frac{m_{\sigma}^*}{n_{\sigma} e^2 \tau_{\sigma}(\Theta,\Phi;\theta,\phi)}, 
\end{eqnarray}
where $e$ is the electron charge and 
$n_\sigma$ ($m^*_\sigma$) is the number density (effective mass)
of the electrons in the conduction band 
of the $\sigma$ spin.\cite{Ibach,Grosso} 
The conduction band consists of 
the s, p, and conductive d states.\cite{Kokado1} 
In addition, $1/\tau_{\sigma}(\Theta,\Phi;\theta,\phi)$ is 
the scattering rate of the conduction electron of the $\sigma$ spin 
at ($\Theta, \Phi$) and ($\theta, \phi$), 
expressed by
\begin{eqnarray}
\label{tau_inv}
&&\frac{1}{\tau_{\sigma}(\Theta,\Phi;\theta,\phi)} 
= \frac{1}{\tau_{s,\sigma}} + 
\sum_{\substack{j=xy, yz, xz\\x^2-y^2, 3z^2-r^2}} \sum_{\varsigma=+, -}
\frac{1}{\tau_{s,\sigma \to j,\varsigma} (\Theta,\Phi;\theta,\phi)}, 
\end{eqnarray}
with
\begin{eqnarray}
\label{tau_sd_inv}
&&\frac{1}{\tau_{s,\sigma \to j,\varsigma}(\Theta,\Phi;\theta,\phi)} = 
\frac{2 \pi}{\hbar} n_{\rm imp} N_{\rm n} {V_{\rm imp}(R_{\rm n})}^2 
\left| (\psi_{j,\varsigma} (\theta,\phi)|
e^{i{\mbox{\boldmath $k$}}_\sigma 
\cdot {\mbox{\boldmath $r$}}},\chi_\sigma (\theta,\phi) \rangle \right|^2 
D_{j,\varsigma}^{(d)}, \nonumber \\\\
&&|e^{i{\mbox{\boldmath $k$}}_\sigma 
\cdot {\mbox{\boldmath $r$}}},\chi_\sigma (\theta,\phi) \rangle=
\frac{1}{\sqrt{\Omega}} e^{i{\mbox{\boldmath $k$}}_\sigma 
\cdot {\mbox{\boldmath $r$}}}\chi_\sigma (\theta,\phi), \\
\label{kkk}
&&{\mbox{\boldmath $k$}}_\sigma
=(k_{x,\sigma}, k_{y,\sigma}, k_{z,\sigma})
=k_\sigma (\sin \Theta \cos \Phi, \sin \Theta \sin \Phi, \cos \Theta), \\
&&{\mbox{\boldmath $r$}}=(x,y,z), 
\end{eqnarray}
where $k_\sigma = |{\mbox{\boldmath $k$}}_\sigma|$. 
Here, 
$1/\tau_{s,\sigma}$ is the $s$--$s$ scattering rate, 
which is considered to be independent of 
($\Theta, \Phi$) and ($\theta, \phi$). 
The $s$--$s$ scattering means that 
the conduction electron of the $\sigma$ spin 
is scattered into the conduction state of the $\sigma$ spin 
by nonmagnetic impurities and phonons.\cite{Kokado4} 
The quantity $1/\tau_{s,\sigma \to j,\varsigma}(\Theta,\Phi;\theta,\phi)$ 
is the $s$--$d$ scattering rate 
at ($\Theta, \Phi$) and ($\theta, \phi$).\cite{Kokado1,Kokado2} 
The $s$--$d$ scattering means that 
the conduction electron of the $\sigma$ spin 
is scattered into 
the $\sigma$ spin state in 
$|\psi_{j,\varsigma} (\theta,\phi))$ of Eq. (\ref{|m,chi_s)}) 
by nonmagnetic impurities. 
The conduction state of the $\sigma$ spin 
$|e^{i{\mbox{\boldmath $k$}}_\sigma 
\cdot {\mbox{\boldmath $r$}}},\chi_\sigma (\theta,\phi) \rangle$ is represented by 
the plane wave, 
where 
${\mbox{\boldmath $k$}}_\sigma$ 
is the Fermi wave vector of the $\sigma$ spin 
at ($\Theta, \Phi$), 
${\mbox{\boldmath $r$}}$ is the position of the conduction electron, 
and $\Omega$ is the volume of the system. 
The quantity 
$V_{\rm imp}(R_{\rm n})$ is 
the scattering potential at $R_{\rm n}$ 
due to a single impurity, 
where $R_{\rm n}$ is the distance between 
the impurity and the nearest-neighbor host atom.\cite{Kokado1} 
The quantity 
$N_{\rm n}$ is the number of nearest-neighbor host atoms 
around a single impurity,\cite{Kokado1} 
$n_{\rm imp}$ is the number density of impurities, 
and 
$\hbar$ is the Planck constant $h$ divided by 2$\pi$. 
The quantity $D_{j,\varsigma}^{(d)}$ represents 
the partial density of states (PDOS) of 
the wave function of the tight-binding model 
for the d state of the $j$ orbital and $\varsigma$ spin 
at 
$E_{\mbox{\tiny F}}$, 
as described in Appendix B in Ref. \citen{Kokado1}.

%We substitute $|\psi_{j,\varsigma} (\theta,\phi))$ 
Substituting $|\psi_{j,\varsigma} (\theta,\phi))$ 
of Eq. (\ref{|m,chi_s)}) 
into 
$1/\tau_{s,\sigma \to j,\varsigma}(\Theta,\Phi;\theta,\phi)$ 
of Eq. (\ref{tau_sd_inv}) 
%Performing the integration with respect to ${\mbox{\boldmath $r$}}$ in $(\psi_{j,\varsigma} (\theta,\phi)|e^{i{\mbox{\boldmath $k$}}_\sigma \cdot {\mbox{\boldmath $r$}}},\chi_\sigma (\theta,\phi) \rangle$, 
we can rewrite 
$\sum_j \sum_\varsigma 
1/\tau_{s,\sigma \to j,\varsigma}(\Theta,\Phi;\theta,\phi)$ 
in Eq. (\ref{tau_inv}) 
as\cite{comment1} 
\begin{eqnarray}
\label{s-d_general}
\sum_{j} \sum_{\varsigma}
\frac{1}{\tau_{s,\sigma \to j,\varsigma} (\Theta,\Phi;\theta,\phi)}
=
\frac{2\pi}{\hbar} 
n_{\rm imp}N_{\rm n}v_\sigma^2 
\sum_{j} \sum_{\varsigma}
P_{\sigma}^{j,\varsigma}(\Theta,\Phi;\theta,\phi)
D_{j, \varsigma}^{(d)},
\end{eqnarray}
with
\begin{eqnarray}
\label{v_sigma}
&&
\hspace*{-0.5cm}
v_{\sigma}=V_{\rm imp}(R_{\rm n}) g_\sigma, \\
\label{g_sigma}
&&
\hspace*{-0.5cm}
g_\sigma = 
- \frac{192 \pi \Gamma \zeta k_\sigma^2 
}{\sqrt{\Omega}(k_\sigma^2 + \zeta^2)^4}, \\
\label{P_{m,sigma}}
&&
\hspace*{-0.5cm}
P_{\sigma}^{j,\varsigma}(\Theta,\Phi;\theta,\phi)
=\left| \varphi_{\sigma}^{j,\varsigma}
(\Theta, \Phi; \theta,\phi) 
\right|^2, \\
\label{dg*}
&&
\hspace*{-0.5cm}
\displaystyle{\varphi_{\sigma}^{j,\varsigma}
(\Theta, \Phi; \theta,\phi) 
= 
c_{xy,\sigma}^{j,\varsigma}(\theta,\phi) x_0 y_0 
+c_{yz,\sigma}^{j,\varsigma}(\theta,\phi) y_0 z_0 
+c_{xz,\sigma}^{j,\varsigma}(\theta,\phi) x_0z_0}  \nonumber \\
&& \hspace*{2.6cm}
\displaystyle{ 
+c_{x^2 -y^2,\sigma}^{j,\varsigma}(\theta,\phi) \frac{1}{2} (x_0^2-y_0^2)
+ c_{3z^2 -r^2,\sigma}^{j,\varsigma}(\theta,\phi) \frac{1}{2 \sqrt{3}} (3z_0^2-r_0^2)}, 
\end{eqnarray}
where 
$x_0 = \sin \Theta \cos \Phi$, 
$y_0=\sin \Theta \sin \Phi$, $z_0=\cos \Theta$, 
and $r_0=1$. 
Here, 
$c_{i,\sigma}^{j,\varsigma}(\theta,\phi)$ 
is numerically obtained 
by applying the exact diagonalization method 
to ${\cal H}$ of Eq. (\ref{Hamiltonian}) [also see Eq. (\ref{|m,chi_s)})]. 
The quantity 
$\varphi_{\sigma}^{j,\varsigma}
(\Theta, \Phi; \theta,\phi)$ 
is proportional to the probability amplitude 
of the d states of the $\sigma$ spin 
of the ${\mbox{\boldmath $I$}}$ direction 
(i.e., the ${\mbox{\boldmath $k$}}_\sigma/k_\sigma$ direction) 
in $|\psi_{j,\varsigma} (\theta,\phi))$ of Eq. (\ref{|m,chi_s)}). 
In addition, 
$P_{\sigma}^{j,\varsigma}(\Theta,\Phi;\theta,\phi)$ 
is proportional to the probability density 
of the d states of the $\sigma$ spin 
of the ${\mbox{\boldmath $I$}}$ direction 
in $|\psi_{j,\varsigma} (\theta,\phi))$ of Eq. (\ref{|m,chi_s)}). 
We emphasize that 
$\left| (\psi_{j,\varsigma} (\theta,\phi)|
e^{i{\mbox{\boldmath $k$}}_\sigma 
\cdot {\mbox{\boldmath $r$}}},\chi_\sigma (\theta,\phi) \rangle \right|^2 $ in Eq. (\ref{tau_sd_inv}) 
is finally expressed by 
$P_{\sigma}^{j,\varsigma}(\Theta,\Phi;\theta,\phi)$.

Using Eqs. (\ref{amr^i(theta,phi)}), 
(\ref{rho^im})$-$(\ref{tau_inv}), 
and 
(\ref{s-d_general})$-$(\ref{dg*}), 
we can calculate 
${\rm AMR}(\Theta,\Phi;\theta,\phi)$ of Eq. (\ref{amr^i(theta,phi)}) 
for various ferromagnets. 
As a check, 
we now 
examine 
${\rm AMR}(\Theta,\Phi;\theta,\phi)$ for a 
system 
with a constant PDOS, 
i.e., $D_{j,\varsigma}^{(d)} \equiv D^{(d)}$, 
where $D^{(d)}$ is the constant. 
In this system, 
$\sum_{j} \sum_{\varsigma}
1/\tau_{s,\sigma \to j,\varsigma} (\Theta,\Phi;\theta,\phi)$ 
of Eq. (\ref{s-d_general}) becomes constant, i.e., 
\begin{eqnarray}
\label{s-d_const}
\sum_{j} \sum_{\varsigma}
\frac{1}{\tau_{s,\sigma \to j,\varsigma} (\Theta,\Phi;\theta,\phi)}
=
\frac{2\pi}{\hbar} 
n_{\rm imp}N_{\rm n}v_\sigma^2 
\frac{D^{(d)}}{3}. 
\end{eqnarray}
Here, Eq. (\ref{s-d_const}) is obtained by using 
the following condition: 
\begin{eqnarray}
\label{p_const}
\sum_{j} \sum_{\varsigma}
P_\sigma^{j,\varsigma} (\Theta,\Phi;\theta,\phi)
=\frac{1}{3},
\end{eqnarray}
which is derived 
from 
Eqs. (\ref{P_{m,sigma}}), (\ref{dg*}), 
and the condition of Eq. (\ref{condition}). 
As a result, 
the system exhibits ${\rm AMR}(\Theta,\Phi;\theta,\phi)=0$.

\section{Application to TAMR effect}
\label{sec_appl}

We apply 
the theory presented in Sect. \ref{sec_theory} to 
the TAMR effect of ${\mbox{\boldmath $I$}}//[100]$ 
and ${\mbox{\boldmath $M$}}$ in the (100) plane 
for ferromagnets. 
This case has $\Theta=\pi/2$, $\Phi=0$, and $\phi=\pi/2$. 
Here, we introduce the following quantities 
for Eqs. (\ref{+spin})$-$(\ref{|m,chi_s)}), 
(\ref{rho_sigma^i}), 
(\ref{tau_sd_inv}), 
and (\ref{P_{m,sigma}}): 
\begin{eqnarray}
\label{no_phi2}
&&\chi_\sigma (\theta,\pi/2) \equiv \chi_\sigma (\theta), \\
\label{no_phi1}
&&|\psi_{j,\varsigma} (\theta,\pi/2)) \equiv |\psi_{j,\varsigma} (\theta)), \\
\label{no_phi3}
&&c_{i,\sigma}^{j,\varsigma}(\theta, \pi/2) \equiv c_{i,\sigma}^{j,\varsigma}(\theta), \\
\label{no_phi31}
&&\rho_\sigma (\pi/2,0;\theta,\pi/2) \equiv \rho_\sigma (\theta), \\
\label{no_phi35}
&&\frac{1}{\tau_{s,\sigma \to j,\varsigma} (\pi/2,0;\theta,\pi/2)} 
\equiv 
\frac{1}{\tau_{s,\sigma \to j,\varsigma} (\theta)}, \\
\label{no_phi4}
&&P_{\sigma}^{j,\varsigma} (\pi/2,0;\theta,\pi/2)
\equiv P_{\sigma}^{j,\varsigma} (\theta). 
\end{eqnarray}

\subsection{Coefficients}
\label{coefficients}
When 
TAMR$(\theta)$ of Eq. (\ref{TAMR}) 
is expressed as Eq. (\ref{Tr-AMR(theta)}), 
we determine $C_0$ to be 
$-C_2-C_4$ from the following relation: 
\begin{eqnarray}
\label{Tr-AMR0}
&&{\rm TAMR}(0) = C_0 + C_2 + C_4 =0. 
\end{eqnarray}
In addition, $C_2$ and $C_4$ are obtained from 
the following equations: 
\begin{eqnarray}
&&{\rm TAMR}(\pi/4) = -C_2 -2 C_4 =f_{\pi/4}, \\
\label{Tr-AMRpi/2}
&&{\rm TAMR}(\pi/2) = - 2C_2  =f_{\pi/2},
\end{eqnarray}
where $f_{\pi/4}$ and $f_{\pi/2}$, respectively, represent 
the numerical values of ${\rm TAMR}(\pi/4)$ and ${\rm TAMR}(\pi/2)$ 
calculated by applying the theory presented in Sect. \ref{sec_theory}. 
The coefficients 
$C_2$ and $C_4$ are therefore expressed as
\begin{eqnarray}
\label{C2_EDM}
&&C_2 = - \frac{1}{2} f_{\pi/2}, \\
\label{C4_EDM}
&&C_4 = \frac{1}{4} f_{\pi/2} - \frac{1}{2} f_{\pi/4}. 
\end{eqnarray}

\subsection{Probability density}

Substituting 
$\Theta=\pi/2$, $\Phi=0$, and $\phi = \pi/2$ 
into Eqs. (\ref{P_{m,sigma}}) and (\ref{dg*}), 
we obtain 
$P_{\sigma}^{j,\varsigma} (\theta)$ of Eq. (\ref{no_phi4}) 
as 
\begin{eqnarray}
\label{P_m}
&&
P_{\sigma}^{j,\varsigma}(\theta)
=\left| \varphi_{\sigma}^{j,\varsigma}(\pi/2,0;\theta,\pi/2) \right|^2,
\end{eqnarray}
with 
\begin{eqnarray}
\label{d_gamma_x}
&&\displaystyle{ 
\varphi_{\sigma}^{j,\varsigma}(\pi/2,0;\theta,\pi/2)
=
\frac{1}{2} c_{x^2-y^2,\sigma}^{j,\varsigma}(\theta) 
- \frac{1}{2 \sqrt{3}} c_{3z^2-r^2,\sigma}^{j,\varsigma}(\theta), 
}
\end{eqnarray}
where Eq. (\ref{no_phi3}) has been used. 
Here, $\varphi_{\sigma}^{j,\varsigma}(\pi/2,0;\theta,\pi/2)$ 
is proportional to 
the probability amplitude of 
the $d\gamma$ states of the $\sigma$ spin 
of the ${\mbox{\boldmath $I$}}$ 
direction ($x$ direction) 
in $|\psi_{j,\varsigma} (\theta))$ of 
Eqs. (\ref{|m,chi_s)}) and (\ref{no_phi1}), 
where 
the $d\gamma$ states of the $\sigma$ spin consist of 
$|x^2-y^2, \chi_\sigma (\theta) \rangle$ and 
$|3z^2-r^2, \chi_\sigma (\theta) \rangle$. 
The $d\gamma$ states of the $\sigma$ spin 
of the ${\mbox{\boldmath $I$}}$ direction ($x$ direction) 
are called 
``$d\gamma_{x,\sigma}$ states'' hereafter. 
%from now on. 
The quantity 
$P_{\sigma}^{j,\varsigma}(\theta)$ 
of Eq. (\ref{P_m}) 
is proportional to the probability density of 
the $d\gamma_{x,\sigma}$ states 
in $|\psi_{j,\varsigma} (\theta))$ of 
Eqs. (\ref{|m,chi_s)}) and (\ref{no_phi1}).

As shown in Eq. (\ref{p_const}), 
$P_{\sigma}^{j,\varsigma} (\theta)$ of 
Eq. (\ref{P_m}) also has the following condition:
\begin{eqnarray}
\label{p_const1}
\sum_{j} \sum_{\varsigma}
P_\sigma^{j,\varsigma} (\theta)
=\frac{1}{3}.
\end{eqnarray}
Equation (\ref{p_const1}) will be used 
in Sects. \ref{cubic_sym} and \ref{tetra_sym}.

\section{TAMR effect for strong ferromagnets}
\label{strong}

We numerically calculate 
$C_2$ of Eq. (\ref{C2_EDM}), $C_4$ of Eq. (\ref{C4_EDM}), 
and ${\rm TAMR}(\theta)$ of Eq. (\ref{TAMR}) 
for strong ferromagnets 
with $D_{j,+}^{(d)} = 0$ and $\sum_j D_{j,-}^{(d)} \ne 0$.\cite{SW_FM} 
We here consider three cases of crystal field: 
%, i.e., 
no crystal field, 
the crystal field of cubic symmetry, 
and 
the crystal field of tetragonal symmetry.\cite{Yosida1} 
In the case of tetragonal symmetry, 
a tetragonal distortion exists in the [001] direction.

\subsection{Systems and parameters}
\label{model_param}

We describe the system and parameters for the calculation.

\begin{figure}[ht]
\begin{center}
\includegraphics[width=0.5\linewidth]{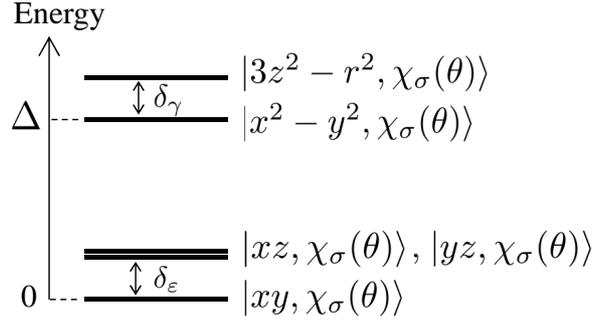}
\caption{
Energy levels of the d states 
in the crystal field of tetragonal symmetry.\cite{Yosida1} 
The second excited states are doubly degenerate. 
The energy levels are measured from 
%$E_{xy}$. 
the energy level of $|xy, \chi_\sigma (\theta) \rangle$. 
The energy levels of the d states 
in the crystal field of cubic symmetry 
correspond to 
the case of $\delta_\varepsilon = \delta_\gamma =0$.\cite{Yosida1} 
}
\label{energy}
\end{center}
\end{figure}

In Fig. \ref{energy}, we show the energy levels 
of the d states 
in the crystal field of tetragonal symmetry.\cite{Yosida1} 
The states 
$|xy, \chi_\sigma (\theta) \rangle$, 
$|yz, \chi_\sigma (\theta) \rangle$, 
and 
$|xz, \chi_\sigma (\theta) \rangle$ 
are 
called $d\varepsilon$ states 
and 
$|x^2-y^2, \chi_\sigma (\theta) \rangle$ 
and 
$|3z^2-r^2, \chi_\sigma (\theta) \rangle$ are 
$d\gamma$ states. 
%We here have $E_{x^2-y^2}-E_{xy} \equiv \Delta$, $E_{xz}-E_{xy} = E_{yz}-E_{xy} \equiv \delta_\varepsilon$, and $E_{3z^2-r^2}-E_{x^2-y^2} \equiv \delta_\gamma$. 
The quantity $\Delta$ is the energy level of $|x^2-y^2, \chi_\sigma (\theta)  \rangle$ measured from that of $|xy, \chi_\sigma (\theta)  \rangle$, 
i.e., 
$\Delta=E_{x^2-y^2}-E_{xy}$. 
The quantity 
$\delta_\varepsilon$ is the energy level of $|xz, \chi_\sigma (\theta)  \rangle$ (or $|yz, \chi_\sigma (\theta)  \rangle$) 
measured from that of $|xy, \chi_\sigma (\theta)  \rangle$, 
i.e., 
$\delta_\varepsilon = E_{xz}-E_{xy} = E_{yz}-E_{xy}$. 
The quantity 
$\delta_\gamma$ is the energy level of $|3z^2-r^2, \chi_\sigma (\theta)  \rangle$ 
measured from that of $|x^2-y^2, \chi_\sigma (\theta)  \rangle$, 
i.e., 
$\delta_\gamma=E_{3z^2-r^2}-E_{x^2-y^2}$. 
Also note that the crystal field of cubic symmetry 
corresponds to 
the case of $\delta_\varepsilon = \delta_\gamma=0$.\cite{Yosida1}

%Following our previous study,\cite{Kokado2} 
In accordance with our previous study,\cite{Kokado2} 
we introduce the following quantities: 
\begin{eqnarray}
\label{r_s-s}
&& r=\frac{\rho_{s,-}}{\rho_{s,+}}, \\
\label{r_s-d}
&& r_{s,\sigma \to j,\varsigma}=
\frac{\rho_{s,\sigma \to j,\varsigma}}{\rho_{s,+}}, 
\end{eqnarray}
with
\begin{eqnarray}
\label{rho_s_pm}
&&\rho_{s,\sigma}= \frac{m_\sigma^*}{n_\sigma e^2 \tau_{s,\sigma}}, \\
\label{rho_m_pm}
&&\rho_{s,\sigma \to j,\varsigma}
= \frac{m_\sigma^*}{n_\sigma e^2 \tau_{s,\sigma \to j,\varsigma}^{(0)}}, 
\end{eqnarray}
where $\rho_{s,\sigma}$ is the $s$--$s$ resistivity 
and 
$\rho_{s,\sigma \to j,\varsigma}$ is the $s$--$d$ resistivity. 
Here, 
$1/\tau_{s,\sigma \to j,\varsigma}^{(0)}$ 
is the $s$--$d$ scattering rate\cite{comment_tau} 
given by
\begin{eqnarray}
\label{1/tau_m_pm}
&&\frac{1}{\tau_{s,\sigma \to j,\varsigma}^{(0)}}=
\frac{2\pi}{\hbar} 
n_{\rm imp}N_{\rm n}\frac{1}{3}v_\sigma^2 D_{j, \varsigma}^{(d)}. 
\end{eqnarray}
As a result, 
$r_{s,\sigma \to j,\varsigma}$ of Eq. (\ref{r_s-d}) 
has 
\begin{eqnarray}
\label{rsd_propto_D}
r_{s,\sigma \to j,\varsigma} \propto D_{j,\varsigma}^{(d)}. 
\end{eqnarray}
In addition to the above-mentioned $D_{j,+}^{(d)}=0$, 
$D_{j,-}^{(d)}$ is set to be 
\begin{eqnarray}
\label{D_xy}
&&D_{xy,-}^{(d)} \equiv D_{\varepsilon 1,-}^{(d)}, \\
&&D_{xz,-}^{(d)} =D_{yz,-}^{(d)} 
\equiv D_{\varepsilon 2,-}^{(d)}, \\
&&D_{x^2-y^2,-}^{(d)} \equiv D_{\gamma 1,-}^{(d)}, \\
\label{D_3z2}
&&D_{3z^2-r^2,-}^{(d)}  
\equiv D_{\gamma 2,-}^{(d)}, 
\end{eqnarray}
on the basis of the energy levels in Fig. \ref{energy}. 
Using Eqs. (\ref{r_s-d})--(\ref{D_3z2}), we then have
\begin{eqnarray}
&&r_{s,\sigma \to xy,-} \equiv 
r_{s,\sigma \to \varepsilon 1,-},\\
&&r_{s,\sigma \to xz,-}=r_{s,\sigma \to yz,-} \equiv 
r_{s,\sigma \to \varepsilon 2,-},\\
&&r_{s,\sigma \to x^2-y^2,-} \equiv r_{s,\sigma \to \gamma 1,-}, \\
&&r_{s,\sigma \to 3z^2-r^2,-} \equiv r_{s,\sigma \to \gamma 2,-}, 
\end{eqnarray}
and also $r_{s,\sigma \to j,+} =0$. 
Actually, there may be a difference in the values 
between $D_{\varepsilon 1,-}^{(d)}$ and $D_{\varepsilon 2,-}^{(d)}$ 
or between $D_{\gamma 1,-}^{(d)}$ and $D_{\gamma 2,-}^{(d)}$, 
as found from the energy levels in Fig. \ref{energy}. 
%\textcolor{red}
%{
In this study, however, 
we ignore such a difference 
under the assumption of 
%assuming that 
%the difference 
$|D_{\varepsilon 1,-}^{(d)} - D_{\varepsilon 2,-}^{(d)}|/D_{\varepsilon 1,-}^{(d)}\ll 1$ and $|D_{\gamma 1,-}^{(d)}-D_{\gamma 2,-}^{(d)}|/D_{\gamma 1,-}^{(d)} \ll 1$, 
%where the assumption 
which 
may be valid for 
%which may be valid for 
%which may be justified for 
%comes from 
%are sufficiently small 
%under 
the parameters with 
%the after-mentioned parameters with 
$\delta \sim \lambda < \Delta \ll H$ mentioned later. 
%below. 
%}
%$\delta < \Delta$. 
%$\delta < \Delta < H$. 
% described later. 
%the present parameters with $\delta < \Delta < H$ described later. 
%in order to simplify discussion through the parameter reduction. 
Namely, we set 
\begin{eqnarray}
\label{DOS_e}
&&D_{\varepsilon 1,-}^{(d)} = D_{\varepsilon 2,-}^{(d)}
\equiv D_{\varepsilon,-}^{(d)}, \\
\label{DOS_g}
&&D_{\gamma 1,-}^{(d)} = D_{\gamma 2,-}^{(d)}
\equiv D_{\gamma,-}^{(d)},
\end{eqnarray}
and 
\begin{eqnarray}
&&r_{s,\sigma \to \varepsilon 1,-}= r_{s,\sigma \to \varepsilon 2,-}
\equiv r_{s,\sigma \to \varepsilon,-}, \\
&&r_{s,\sigma \to \gamma 1,-} = r_{s,\sigma \to \gamma 2,-}
\equiv r_{s,\sigma \to \gamma,-}. 
\end{eqnarray}
%\textcolor{red}
%{
In addition, 
%in accordance with 
in a conventional manner,\cite{comment_sigma} 
%in a conventional manner,\cite{Berger,Kokado1,Kokado3}
%following our previous study,\cite{Kokado3} 
%for simplicity, we put 
%we consider a simple case with 
%we consider a simple case with 
we consider a simple system with 
%we adopt a simple system with 
%we adopt a simple system with 
%ignore the $\sigma$ dependence of 
%$r_{s,\sigma \to j,\varsigma}$ on the assumption of 
%we assume 
$n_+=n_-$, 
%$n_\uparrow$=$n_\downarrow$, 
$m_+^*=m_-^*$, and $v_+=v_-$, 
where 
$v_+=v_-$ is satisfied by setting $k_+=k_-$ 
in Eqs. (\ref{v_sigma}) and (\ref{g_sigma}). 
In this system, 
$r_{s,\sigma \to j,-}$ of Eq. (\ref{r_s-d}) 
%$r_{s,\sigma \to j,\varsigma}$ of Eq. (\ref{r_s-d}) 
is independent of $\sigma$. 
%}
We also introduce the following quantities: 
\begin{eqnarray}
\label{P_varepsilon}
&&P_\sigma^{d\varepsilon,-}(\theta)=P_\sigma^{xy,-} (\theta)+P_\sigma^{yz,-} (\theta)+P_\sigma^{xz,-} (\theta), \\
\label{P_gamma}
&&P_\sigma^{d\gamma,-}(\theta)=P_\sigma^{x^2-y^2,-} (\theta)+P_\sigma^{3z^2-r^2,-} (\theta),
\end{eqnarray}
where $P_{\sigma}^{j,\varsigma}(\theta)$ has been given by Eq. (\ref{P_m}). 
The quantity $P_\sigma^{d\varepsilon,-}(\theta)$ 
is proportional to 
the sum of 
the probability densities of the $d\gamma_{x,\sigma}$ states 
in $|\psi_{xy,-}(\theta))$, 
$|\psi_{yz,-}(\theta))$, 
and 
$|\psi_{xz,-}(\theta))$. 
The quantity $P_\sigma^{d\gamma,-}(\theta)$ 
is proportional to 
the sum of 
the probability densities of the $d\gamma_{x,\sigma}$ states 
in $|\psi_{x^2-y^2,-}(\theta))$ 
and 
$|\psi_{3z^2-r^2,-}(\theta))$.

As common parameters, we set 
$H=1$ eV, 
$\lambda=\lambda'=0.05$ eV, 
and $r=0.001$ 
bearing a typical strong ferromagnet in mind.\cite{comment_param} 
%\textcolor{red}
%{
In addition, 
the system with no crystal field has 
$\Delta=\delta_\varepsilon=\delta_\gamma=0$ 
and the identical $D_{j,-}^{(d)}$ 
for $j=1$, 2, 3, 4, and 5. 
%$j=1$, 2, 3, $\cdots$, 10. 
%and 
On the basis of the identical $D_{j,-}^{(d)}$ 
and our previous study,\cite{comment_param} 
$r_{s,\sigma \to j,-}$ of Eqs. (\ref{r_s-d}) and (\ref{rsd_propto_D}) 
is set to be 0.01 
% for 
%every $j$. 
%where 
for $j=1$, 2, 3, 4, and 5. 
%}
%for $j=1$, 2, 3, $\cdots$, 10. 
%$j=1$$-$10. 
%$D_{\varepsilon,-}^{(d)}=D_{\gamma,-}^{(d)}$. 
The system with the crystal field of cubic symmetry has 
$\Delta=0.1$ eV and $\delta_\varepsilon=\delta_\gamma=0$.\cite{comment_param} 
For the system with the crystal field of tetragonal symmetry, 
we have $\Delta=0.1$ eV 
and, 
for simplicity, set
%put
\begin{eqnarray}
\label{delta}
&&\delta_\varepsilon=\delta_\gamma \equiv \delta. 
\end{eqnarray}
The value of $\delta$ is assumed to be the same as that of $\lambda$, 
i.e., 0.05 eV. 
We also consider three models for the cubic or tetragonal system: 
%, i.e., 
the $d\varepsilon$, $d\gamma$, and $d\varepsilon+d\gamma$ models. 
The $d\varepsilon$ model has 
$r_{s,\sigma \to \varepsilon,-}=0.01$ (Ref. \citen{comment_param}) 
and 
$r_{s,\sigma \to \gamma,-}=0$, 
which indicate 
$D_{\varepsilon,-}^{(d)} \ne 0$ 
and $D_{\gamma,-}^{(d)}=0$, respectively. 
The $d\gamma$ model has 
$r_{s,\sigma \to \varepsilon,-}=0$ 
and 
$r_{s,\sigma \to \gamma,-}=0.01$, 
which indicate 
$D_{\varepsilon,-}^{(d)} = 0$ 
and $D_{\gamma,-}^{(d)} \ne 0$, respectively. 
The $d\varepsilon+d\gamma$ model has the following three types. 
The first type is 
$r_{s,\sigma \to \gamma,-}=0.01$ 
and 
$r_{s,\sigma \to \varepsilon,-}=0.005$, 
i.e., 
$D_{\varepsilon,-}^{(d)}/D_{\gamma,-}^{(d)}=0.5$. 
The second type is 
$r_{s,\sigma \to \gamma,-}=r_{s,\sigma \to \varepsilon,-}=0.01$, 
i.e., 
$D_{\varepsilon,-}^{(d)}/D_{\gamma,-}^{(d)}=1$. 
The third type is 
$r_{s,\sigma \to \gamma,-}=0.01$ 
and 
$r_{s,\sigma \to \varepsilon,-}=0.015$, 
i.e., 
$D_{\varepsilon,-}^{(d)}/D_{\gamma,-}^{(d)}=1.5$.

\subsection{Calculation result and consideration}
\label{no_crystal}

We find that 
the cubic system 
exhibits the fourfold symmetric TAMR effect, 
while  
the tetragonal system 
shows the twofold and fourfold symmetric TAMR effect.

\subsubsection{No crystal field}
\label{no_crystal}

The system with no crystal field of $\Delta=\delta=0$ 
exhibits $C_2 = C_4 = 0$ and ${\rm TAMR}(\theta)=0$. 
This result 
agrees with that of 
our previous study 
based on the perturbation theory.\cite{comment_sd}

\subsubsection{Crystal field of cubic symmetry}
\label{cubic_sym}

\begin{figure}[ht]
\begin{center}
\includegraphics[width=.5\linewidth]{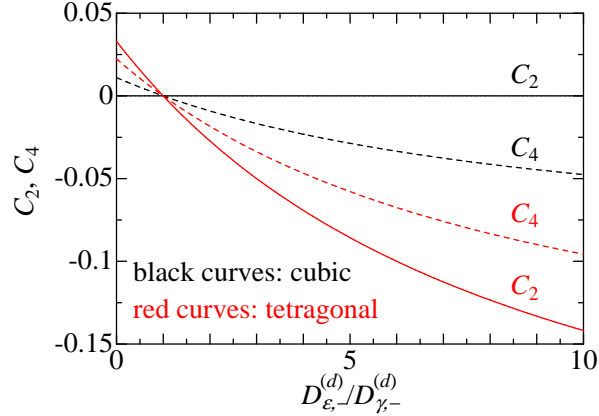} 
\caption{
(Color online) 
The quantity 
$D_{\varepsilon,-}^{(d)}/D_{\gamma,-}^{(d)}$ 
($=r_{s,\sigma \to \varepsilon,-}/r_{s,\sigma \to \gamma,-}$) 
dependences of $C_2$ and $C_4$ 
for a strong ferromagnet 
with a crystal field of cubic symmetry 
or tetragonal symmetry. 
The system has 
$D_{j,+}^{(d)} = 0$ (i.e., $r_{s,\sigma \to j,+}=0$), 
$r_{s,\sigma \to \gamma,-}=0.01$, 
$r=0.001$, 
$H=1$ eV, 
$\lambda=\lambda'=0.05$ eV, 
and $\Delta =0.1$ eV. 
In addition, 
the system with a crystal field of cubic (tetragonal) symmetry 
has 
$\delta =0$ eV ($\delta=0.05$ eV). 
The result for the cubic (tetragonal) system 
is indicated by the black (red) curves. 
The coefficient $C_2$ is represented by the solid line or curve, 
and $C_4$ is shown by the dashed curve. 
%The coefficient $C_2$ ($C_4$) is represented by the solid (dashed) curve. 
}
\label{c_coeff_D}
\end{center}
\end{figure}

We consider 
the system with the crystal field of cubic symmetry. 
In Fig. \ref{c_coeff_D}, 
we show 
the 
$D_{\varepsilon,-}^{(d)}/D_{\gamma,-}^{(d)}$ 
($=r_{s,\sigma \to \varepsilon,-}/r_{s,\sigma \to \gamma,-}$) 
dependence of 
$C_2$ of Eq. (\ref{C2_EDM}) [$C_4$ of Eq. (\ref{C4_EDM})] 
for the system 
with $r_{s,+ \to \gamma,-}=r_{s,- \to \gamma,-}=0.01$ 
by the solid black line [dashed black curve]. 
We find that 
the system has 
$C_2=0$ 
regardless of $D_{\varepsilon,-}^{(d)}/D_{\gamma,-}^{(d)}$. 
In addition, the system shows 
$C_4>0$ for $D_{\varepsilon,-}^{(d)}/D_{\gamma,-}^{(d)} < 1$, 
$C_4=0$ at $D_{\varepsilon,-}^{(d)}/D_{\gamma,-}^{(d)}=1$, 
and $C_4<0$ for $D_{\varepsilon,-}^{(d)}/D_{\gamma,-}^{(d)} >1$. 
By using $C_2=0$ and Eq. (\ref{Tr-AMR0}), 
TAMR$(\theta)$ of Eq. (\ref{Tr-AMR(theta)}) 
is 
rewritten as
\begin{eqnarray}
\label{Tr-AMR_cubic}
{\rm TAMR}(\theta) = C_4 (-1+  \cos 4 \theta). 
\end{eqnarray}
The system therefore exhibits 
the fourfold symmetric TAMR effect with 
${\rm TAMR}(\theta) \le 0$ 
for $D_{\varepsilon,-}^{(d)}/D_{\gamma,-}^{(d)} < 1$ 
and 
${\rm TAMR}(\theta) \ge 0$ 
for $D_{\varepsilon,-}^{(d)}/D_{\gamma,-}^{(d)} > 1$, 
while 
the system has 
${\rm TAMR}(\theta)= 0$ at $D_{\varepsilon,-}^{(d)}/D_{\gamma,-}^{(d)}= 1$.

\begin{figure}[ht]
\begin{center}
\includegraphics[width=.4\linewidth]{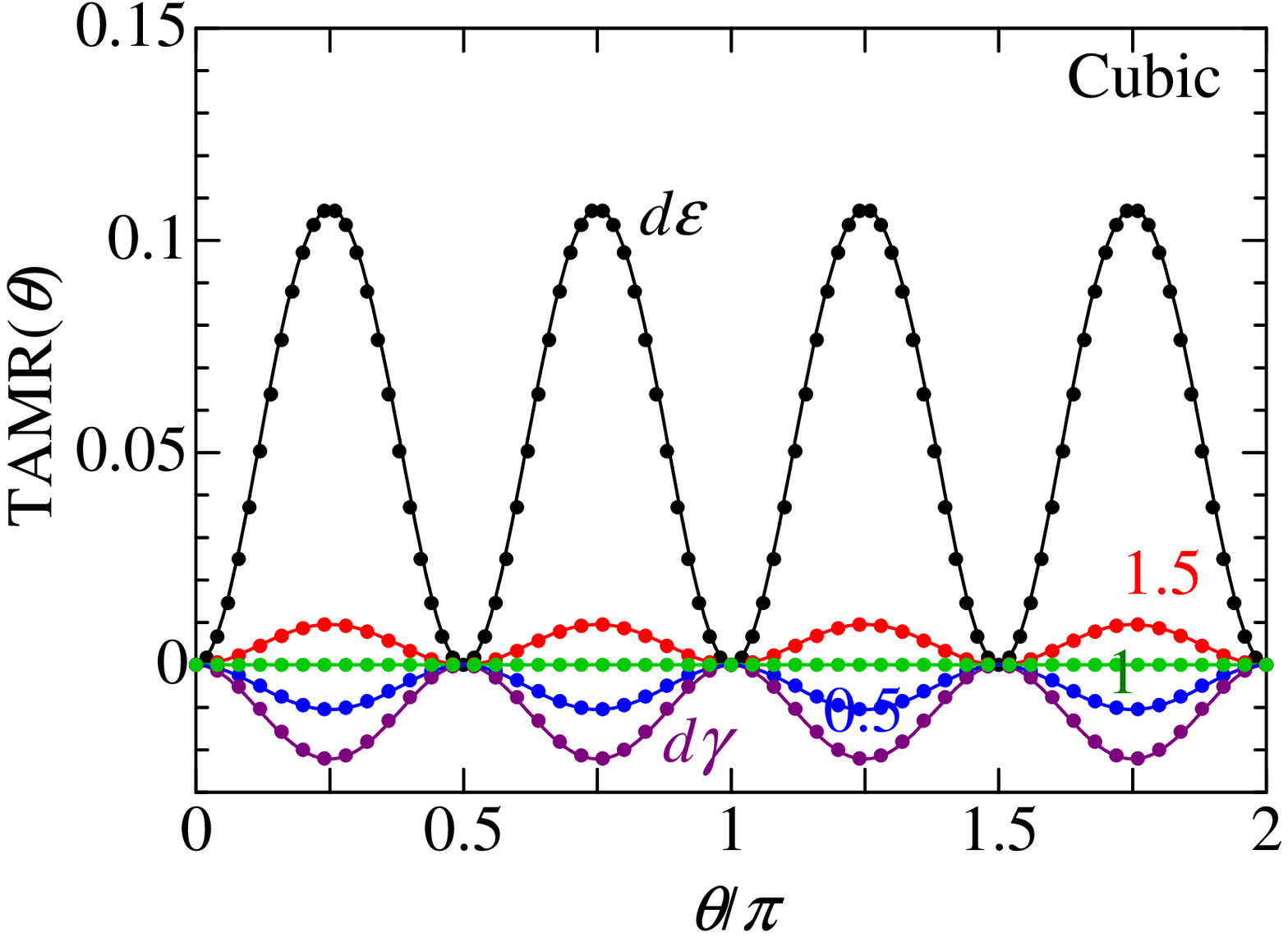} \\[-0.8cm]
\hspace*{-7cm}(a)\\
\vspace{0.3cm}
\includegraphics[width=.4\linewidth]{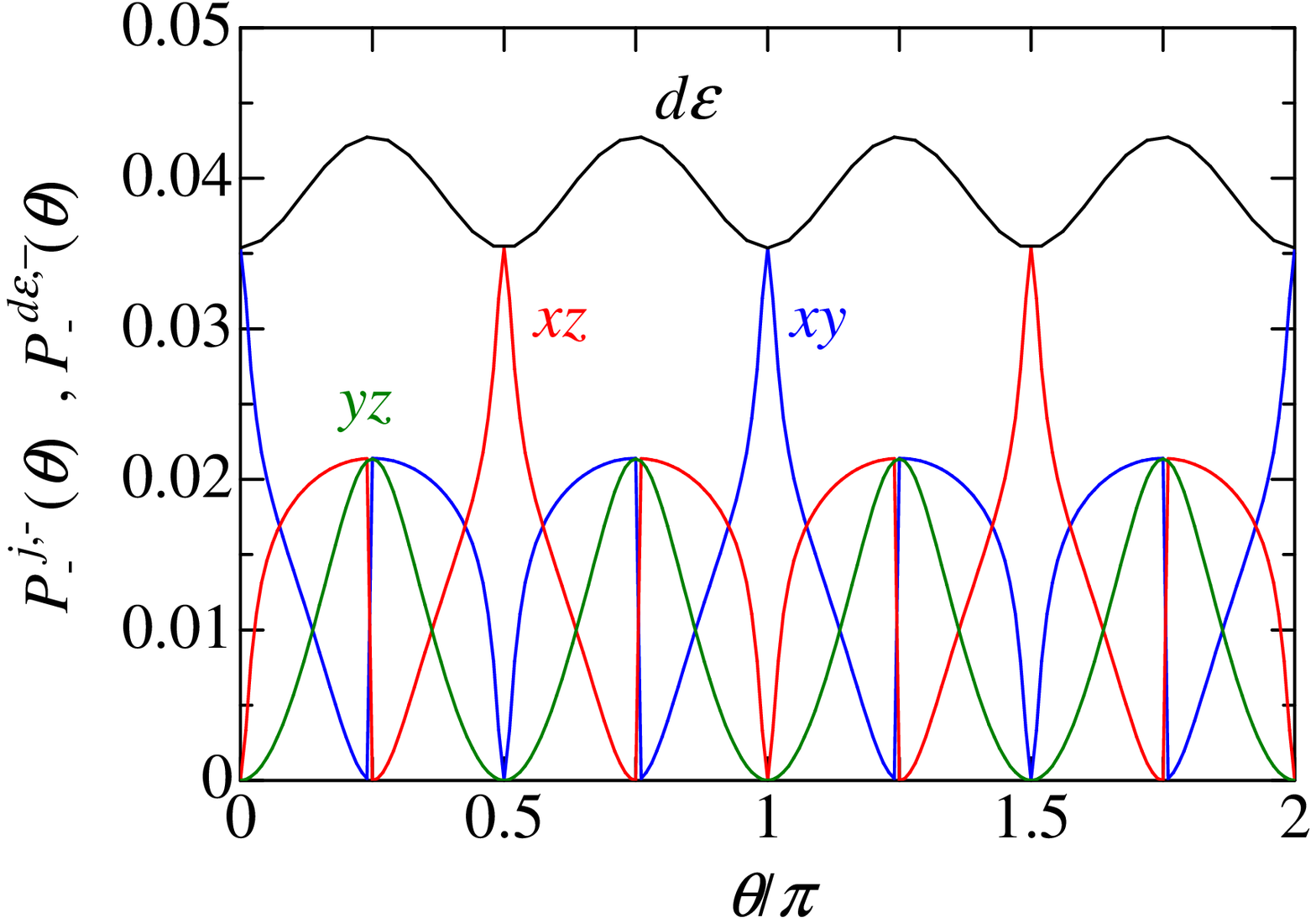} \\[-0.8cm]
\hspace*{-7cm}(b)\\
\vspace{0.3cm}
\includegraphics[width=.4\linewidth]{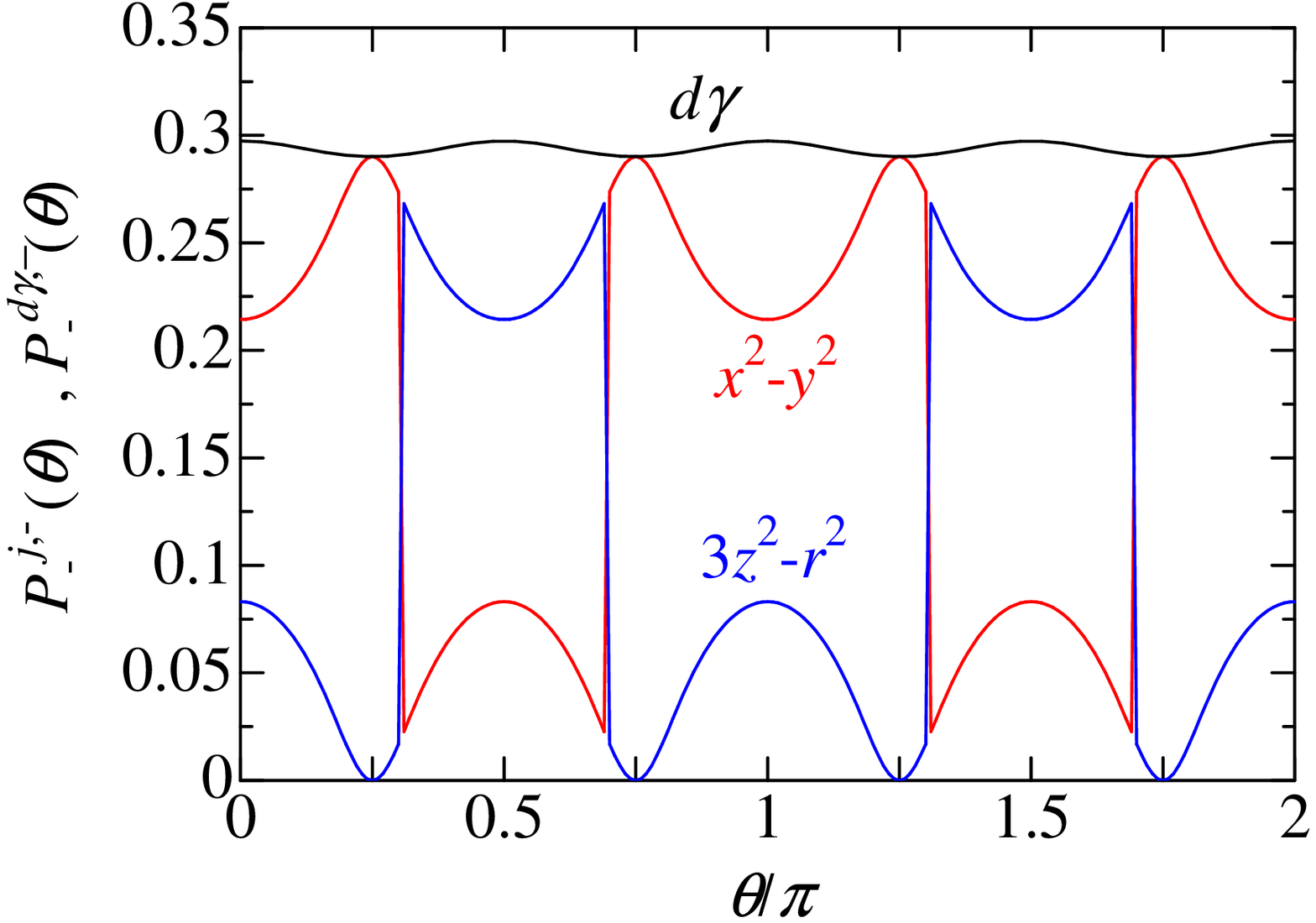} \\[-0.8cm]
\hspace*{-7cm}(c)\\
\vspace{0.3cm}
\includegraphics[width=.4\linewidth]{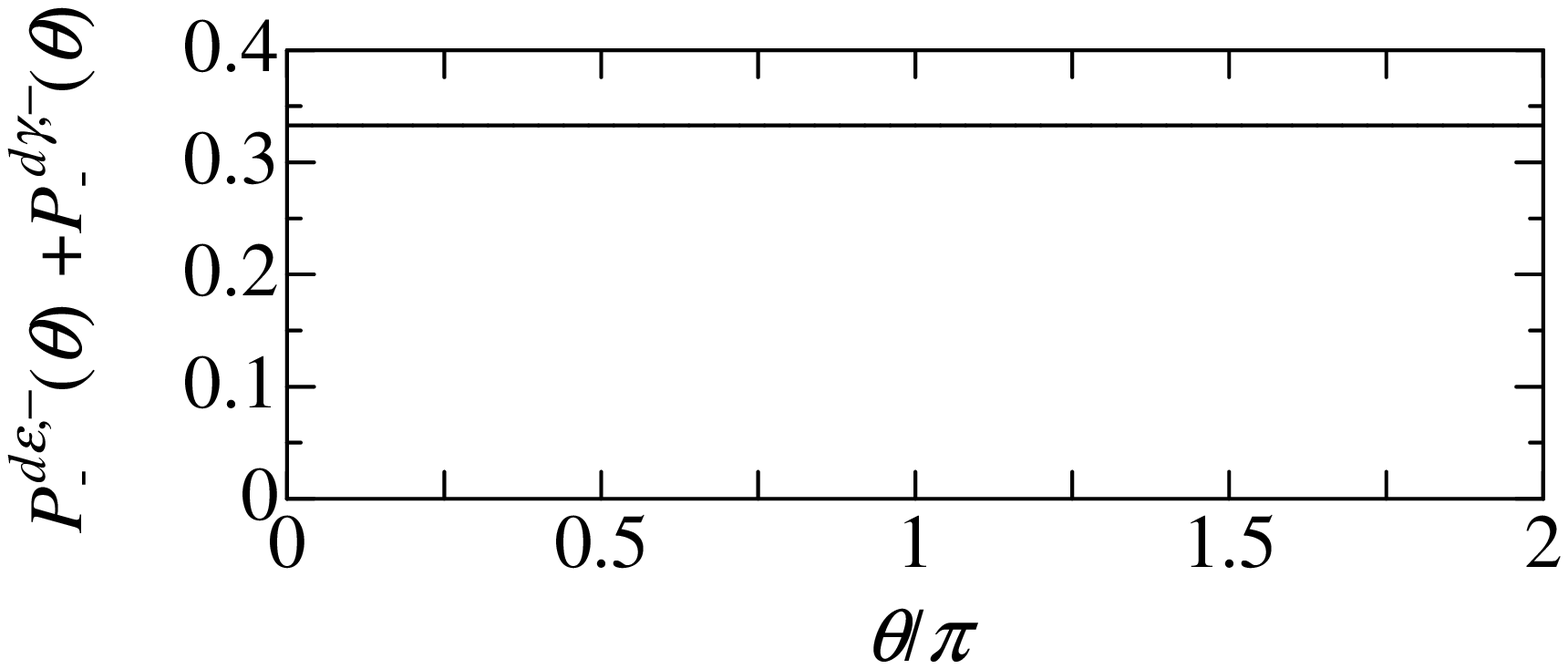} \\[-0.8cm]
\hspace*{-7cm}(d)\\
\vspace{0.3cm}
\caption{
(Color) 
The case of strong ferromagnets 
with a crystal field of cubic symmetry. 
The system has 
$D_{j,+}^{(d)} = 0$ (i.e., $r_{s,\sigma \to j,+}=0$), 
$r=0.001$, 
$H=1$ eV, 
$\lambda=\lambda'=0.05$ eV, 
and $\Delta =0.1$ eV. 
(a) The angle $\theta$ dependences of TAMR$(\theta)$ 
for the $d\varepsilon$, 
$d\gamma$, 
and 
$d\varepsilon+d\gamma$ models. 
The results for the $d\varepsilon$ and $d\gamma$ models 
are indicated by the black and purple curves, respectively. 
The results for the $d\varepsilon+d\gamma$ model 
with $D_{\varepsilon,-}^{(d)}/D_{\gamma,-}^{(d)} = 0.5$, 1, and 1.5 
are indicated by the blue, green, and red curves, respectively. 
In addition, 
the respective colored dots 
represent 
TAMR$(\theta)$ 
of Eq. (\ref{Tr-AMR_cubic}) 
with 
the evaluated $C_4$s of Eq. (\ref{C4_EDM}), 
where $C_4$ for each model is 
%summarized 
given in Table \ref{tab_C4}. 
(b) The angle $\theta$ dependences of 
$P_-^{d\varepsilon,-}(\theta)$, 
$P_-^{xy,-}(\theta)$, 
$P_-^{yz,-}(\theta)$, 
and 
$P_-^{xz,-}(\theta)$ for the $d\varepsilon$ model. 
Here, 
$P_-^{d\varepsilon,-}(\theta)$, 
$P_-^{xy,-}(\theta)$, 
$P_-^{yz,-}(\theta)$, 
and 
$P_-^{xz,-}(\theta)$ 
are indicated by the black, blue, green, and red curves, respectively. 
(c) The angle $\theta$ dependences of 
$P_-^{d\gamma,-}(\theta)$, 
$P_-^{x^2 - y^2,-}(\theta)$, 
and 
$P_-^{3z^2 - r^2,-}(\theta)$ 
for the $d\gamma$ model. 
Here, 
$P_-^{d\gamma,-}(\theta)$, 
$P_-^{x^2 - y^2,-}(\theta)$, 
and 
$P_-^{3z^2 - r^2,-}(\theta)$ 
are indicated by the black, red, and blue curves, respectively. 
(d) The angle $\theta$ dependences of 
$P_-^{d\varepsilon,-}(\theta)+ P_-^{d\gamma,-}(\theta)$. 
Here, 
$P_-^{d\varepsilon,-}(\theta) + P_-^{d\gamma,-}(\theta)$ 
is the sum of 
$P_-^{d\varepsilon,-}(\theta)$ in (b) 
and $P_-^{d\gamma,-}(\theta)$ in (c). 
}
\label{tr-amr_c}
\end{center}
\end{figure}

In Fig. \ref{tr-amr_c}(a), we show 
the $\theta$ dependence of TAMR($\theta$) of Eq. (\ref{TAMR}) 
for the $d\varepsilon$, $d\gamma$, and $d\varepsilon+d\gamma$ models. 
The curves represent 
results calculated directly by 
%using 
the exact diagonalization method. 
The dots show 
results obtained by substituting the evaluated $C_4$s 
of Eq. (\ref{C4_EDM}) 
into 
TAMR$(\theta)$ 
of Eq. (\ref{Tr-AMR_cubic}), 
where $C_4$ 
for each model 
is 
%summarized 
given in Table \ref{tab_C4}.
We find 
contradictory behavior of TAMR($\theta$) 
between 
the $d \varepsilon$ model 
and 
$d \gamma$ model. 
Namely, the $d \varepsilon$ model 
exhibits ${\rm TAMR}(\theta)\ge 0$ 
with the highest value 0.11 at 
$\displaystyle{ \theta = \frac{\pi}{4} + \frac{n\pi}{2}}$ 
and 
the lowest value 0 at 
$\displaystyle{\theta =\frac{n\pi}{2}}$, where $n=0, \pm1, \pm2, \cdots$. 
In contrast, 
the $d \gamma$ model 
shows ${\rm TAMR}(\theta)\le 0$ 
with 
the highest value 0 at 
$\displaystyle{\theta =\frac{n\pi}{2}}$ 
and 
the lowest value $-$0.022 at 
$\displaystyle{ \theta = \frac{\pi}{4} + \frac{n\pi}{2}}$, 
where $n=0, \pm1, \pm2, \cdots$. 
The $d \varepsilon+d\gamma$ model 
takes TAMR($\theta$) between 
TAMR($\theta$) of the $d \varepsilon$ model and that of the $d\gamma$ model. 
Namely, the $d \varepsilon+d\gamma$ model has 
$0 \le {\rm TAMR(\theta)} \le 0.0095$ 
for $D_{\varepsilon,-}^{(d)}/D_{\gamma,-}^{(d)}=1.5$, 
${\rm TAMR(\theta)} = 0$ 
at $D_{\varepsilon,-}^{(d)}/D_{\gamma,-}^{(d)}=1$, 
and 
$-0.011 \le {\rm TAMR(\theta)} \le 0$ 
for $D_{\varepsilon,-}^{(d)}/D_{\gamma,-}^{(d)}=0.5$. 
Roughly speaking, 
the positive and large TAMR($\theta$) for the $d \varepsilon$ model 
is reduced 
with the addition of the negative TAMR($\theta$) for the $d\gamma$ model 
[see Eq. (\ref{TAMR_de+dg2})].

\begin{table}[ht]
\begin{center}
\caption{
The coefficient $C_4$ of Eq. (\ref{C4_EDM}) 
for strong ferromagnets 
with a crystal field of cubic symmetry. 
The system has 
$D_{j,+}^{(d)} = 0$ (i.e., $r_{s,\sigma \to j,+}=0$), 
$r=0.001$, 
$H=1$ eV, 
$\lambda=\lambda'=0.05$ eV, 
and $\Delta =0.1$ eV. 
For these systems, we consider 
$d \varepsilon$, $d\gamma$, and $d\varepsilon+d \gamma$ models. 
In particular, 
the $d\varepsilon+d \gamma$ model has 
$D_{\varepsilon,-}^{(d)}/D_{\gamma,-}^{(d)}=0.5$, 1, and 1.5. 
Note that $C_2$ of Eq. (\ref{C2_EDM}) is evaluated to be 0. 
Also, 
TAMR($\theta$) of Eq. (\ref{Tr-AMR_cubic}) 
with each $C_4$ in this table 
is indicated by the dots in Fig. \ref{tr-amr_c}(a). 
}
\begin{tabular}{lc}
\hline 
Model & $C_4$ \\
\hline 
$d\varepsilon$ & $-$0.054\\
$d\gamma$ &  0.011  \\
$d\varepsilon+d \gamma$ ($D_{\varepsilon,-}^{(d)}/D_{\gamma,-}^{(d)}=0.5$) &  0.0053 \\
$d\varepsilon+d \gamma$ ($D_{\varepsilon,-}^{(d)}/D_{\gamma,-}^{(d)}=1$) &  0 \\
$d\varepsilon+d \gamma$ ($D_{\varepsilon,-}^{(d)}/D_{\gamma,-}^{(d)}=1.5$) &  $-$0.0048 \\
\hline
\end{tabular}
\label{tab_C4}
\end{center}
\end{table}

The $\theta$ dependences 
of TAMR($\theta$) for the $d\varepsilon$ and $d\gamma$ models 
directly reflect 
those of $P_-^{d\varepsilon,-}(\theta)$ and 
$P_-^{d\gamma,-}(\theta)$, 
respectively, 
as found from Eqs. (\ref{TAMR_approx})--(\ref{1/tau_approx}), 
(\ref{TAMR_depsilon}), and (\ref{TAMR_dgamma}). 
In particular, 
the contradictory behavior 
of TAMR($\theta$) 
between the $d\varepsilon$ model and the $d\gamma$ model 
arises under the condition of 
$P_-^{d\varepsilon,-}(\theta)+ P_-^{d\gamma,-}(\theta) \approx 1/3$ 
of Eq. (\ref{1/3}). 
In Fig. \ref{tr-amr_c}(b) [(c)], 
we first show the $\theta$ dependences of 
$P_-^{d\varepsilon,-}(\theta)$ [$P_-^{d\gamma,-}(\theta)$] 
by the black curve. 
We find contradictory behavior between 
$P_-^{d\varepsilon,-}(\theta)$ and $P_-^{d\gamma,-}(\theta)$, 
in which 
$P_-^{d\gamma,-}(\theta)$ decreases (increases) 
as $P_-^{d\varepsilon,-}(\theta)$ increases (decreases). 
Next, the $\theta$ dependences of 
$P_-^{d\varepsilon,-}(\theta) + P_-^{d\gamma,-}(\theta)$ 
are shown in Fig. \ref{tr-amr_c}(d). 
We confirm 
that 
$P_-^{d\varepsilon,-}(\theta)+ P_-^{d\gamma,-}(\theta)$ 
takes the constant value, 1/3; 
that is, 
the condition of Eq. (\ref{1/3}) 
is satisfied.

The value of TAMR($\theta$) 
for the $d\varepsilon + d\gamma$ model 
is checked on the basis of 
the approximate expression of TAMR($\theta$) of 
Eqs. (\ref{TAMR_depsilon})--(\ref{TAMR_de+dg2}). 
%Eq. (\ref{TAMR_de+dg}) or (\ref{TAMR_de+dg1}). 
As shown in Eq. (\ref{relation_value}), 
%when $D_{\varepsilon,-}^{(d)}/D_{\gamma,-}^{(d)}=1.5$ [0.5], 
%the value of 
TAMR($\theta$) of Eq. (\ref{TAMR_de+dg2}) 
%[(\ref{TAMR_de+dg1})] 
is less than or equal to 
that for the $d\varepsilon$ model of Eq. (\ref{TAMR_depsilon}) 
and 
greater than or equal to 
%is  than or equal to 
that for the $d\gamma$ model of Eq. (\ref{TAMR_dgamma}). 
%[(\ref{TAMR_dgamma})]. 
In particular, 
%In addition, 
when 
$D_{\varepsilon,-}^{(d)}/D_{\gamma,-}^{(d)}=1$, 
TAMR($\theta$) of Eq. (\ref{TAMR_de+dg}) or (\ref{TAMR_de+dg1}) 
becomes 0 [see Eq. (\ref{0})].

We describe the details of $P_-^{d\varepsilon,-}(\theta)$ 
of Eq. (\ref{P_varepsilon}) 
for the $d\varepsilon$ model [see Fig. \ref{tr-amr_c}(b)]. 
The quantity $P_-^{d\varepsilon,-}(\theta)$ has 
peaks at $\theta = \pi/4 + n\pi/2$, 
with $n=\pm0, \pm1, \pm2, \cdots$. 
The peaks come from 
$P_-^{xy,-}(\pi/4 + n\pi/2)$, 
$P_-^{yz,-}(\pi/4 + n\pi/2)$, 
and $P_-^{xz,-}(\pi/4 + n\pi/2)$. 
Namely, 
the $d \varepsilon$ states consisting of 
$|xy, \chi_- (\theta) \rangle$, 
$|yz, \chi_- (\theta) \rangle$, 
and 
$|xz, \chi_- (\theta) \rangle$ 
are strongly hybridized to the $d\gamma_{x,-}$ states 
at $\theta = \pi/4 + n\pi/2$. 
On the other hand, 
$P_-^{d\varepsilon,-}(n \pi)$ 
[$P_-^{d\varepsilon,-}(\pi/2 + n \pi)$] 
is formed by 
only $P_-^{xy,-}(n \pi)$ [$P_-^{xz,-}(\pi/2 + n \pi)$], 
where $n=0, \pm1, \pm2, \cdots$. 
In short, 
only $|xy, \chi_- (\theta) \rangle$ 
[$|xz, \chi_- (\theta) \rangle$] in the $d\varepsilon$ states 
is 
hybridized to the $d\gamma_{x,-}$ states 
at $\theta = n \pi$ [$\pi/2 + n \pi$]. 
In fact, when the magnetic quantum number is represented by $m$, 
$|xy, \chi_- (n \pi) \rangle$ 
%consisting of the states of 
with $m=-2$ and 2 (Ref. \citen{3d}) 
is coupled to 
$|x^2-y^2, \chi_- (n \pi) \rangle$ 
%consisting of the states of 
with $m=-2$ and 2 
through 
the $\lambda' L_zS_z$ term, 
as found from 
% Table \ref{matrix}, i.e., 
$\langle x^2-y^2,\chi_- (n \pi) | 
{\cal H} | xy, \chi_- (n \pi) \rangle = i \lambda' (-1)^n$ 
(see ${\cal H}$ of Table \ref{matrix}). 
In addition, 
$|xz, \chi_- (\pi/2 + n \pi) \rangle$ 
with $m=-1$ and 1 
is coupled to 
$|x^2-y^2,\chi_- (\pi/2 + n \pi) \rangle$ 
with $m=-2$ and 2 
and 
$|3z^2-r^2,\chi_- (\pi/2 + n \pi) \rangle$ 
%the $d\gamma$ states 
with $m=0$ 
through 
the $\lambda (L_xS_x + L_y S_y)$ term (the so-called mixing term), 
as seen from 
$\langle x^2-y^2,\chi_- (\pi/2 + n \pi)|{\cal H} |xz, \chi_- (\pi/2 + n \pi) \rangle=-i (\lambda/2) (-1)^n$ 
and 
$\langle 3z^2-r^2,\chi_- (\pi/2 + n \pi)|{\cal H} | xz, \chi_- (\pi/2 + n \pi)\rangle=i (\sqrt{3}\lambda/2) (-1)^n$. 
%, respectively. 
%the matrix elements of ${\cal H}$ 
%between 
%$\langle xz, \chi_- (\pi/2 + n \pi)|$ 
%and the $d\gamma$ states. 
%$\langle x^2-y^2,\chi_- (\theta,\pi/2) | {\cal H} | xy, \chi_- (\theta,\pi/2) \rangle = i \lambda' \cos \theta$. 
%at $\theta = n \pi$ [$\pi/2 + n \pi$]. 
As for the respective maximum values of $P_-^{j,-}(\theta)$, 
we find 
$P_-^{xz,-}(\pi/2 + n \pi)=P_-^{xy,-}(n \pi) > P_-^{yz,-}(\pi/4 + n \pi/2)$. 
This relation is attributed to 
the probability density of the $d\gamma_{x,-}$ states 
for the cubic system 
[see Eqs. (\ref{P_m}) and (\ref{d_gamma_x})]. 
Namely, in the case of the cubic system, 
the probability density of the $d\gamma_{x,-}$ states in 
$|xz, \chi_- (\pi/2 + n \pi) )$ 
is the same as that in $|xy, \chi_- (n \pi) )$, 
whereas 
the probability density of the $d\gamma_{x,-}$ states in 
$|yz, \chi_- (\pi/4 + n \pi/2) )$
is smaller than 
that in 
$|xy, \chi_- (n \pi) )$.

We give the details of $P_-^{d\gamma,-}(\theta)$ 
of Eq. (\ref{P_gamma}) 
for the $d\gamma$ model [see Fig. \ref{tr-amr_c}(c)]. 
The quantity $P_-^{d\gamma,-}(\theta)$ has peaks 
at $\theta =n\pi/2$, 
with $n=0, \pm1, \pm2, \cdots$. 
The peaks come from 
$P_-^{x^2-y^2,-}(n\pi/2)$ and $P_-^{3z^2 - r^2,-}(n\pi/2)$. 
Namely, 
the $d \gamma$ states consisting of 
$|\psi_{x^2-y^2, -} (\theta) )$ and 
$|\psi_{3z^2-r^2, -} (\theta) )$ 
include the $d \gamma_{x,-}$ states 
with the highest probability density 
at 
$\theta =n\pi/2$. 
In addition, we find the characteristic relation of 
$P_-^{3z^2 - r^2,-}(\pi/2 + n\pi) = P_-^{x^2-y^2,-}( n\pi)$, 
even though 
the coefficient 
``$1/(2\sqrt{3})$''
of $c_{3z^2-r^2,-}^{3z^2-r^2,-}(\pi/2 +n\pi)$ 
in $P_{-}^{3z^2-r^2,-}(\pi/2 +n\pi)$ 
of Eqs. (\ref{P_m}) and (\ref{d_gamma_x}) 
%[see Eqs. (\ref{P_m}) and (\ref{d_gamma_x})] 
is smaller than 
the coefficient 
``$1/2$'' 
of $c_{x^2-y^2,-}^{x^2-y^2,-}( n\pi)$ 
in $P_{-}^{x^2-y^2,-}( n\pi)$. 
%This relation indicates that 
%$|c_{3z^2-r^2,-}^{x^2-y^2,-}(n\pi)| \ll |c_{x^2-y^2,-}^{x^2-y^2,-}(n\pi)| \sim 1$ and 
%$|c_{3z^2-r^2,-}^{3z^2-r^2,-}(\pi/2 +n\pi)| \sim |c_{x^2-y^2,-}^{x^2-y^2,-}(n\pi)| \sim 1$ does not hold 
%is not realized 
%because of large hybridization between the degenerate $d\gamma$ states. 
This relation 
%comes from the fact 
reflects that 
$|c_{3z^2-r^2,-}^{3z^2-r^2,-}(\pi/2 +n\pi)| 
%\sim |c_{x^2-y^2,-}^{x^2-y^2,-}(n\pi)| 
\sim 1$ 
does not hold 
%is not realized 
because of 
%due to 
large hybridization 
at $\theta = \pi/2 + n \pi$ 
between the degenerate $d\gamma$ states 
through the $d\varepsilon$ states, 
and then, 
$|c_{x^2-y^2,-}^{3z^2-r^2,-}(\pi/2 +n\pi)|$ has non-negligible magnitude 
%[also see Table \ref{matrix}, i.e., 
%as seen from 
[also see 
$\langle 3z^2-r^2,\chi_- (\pi/2 + n \pi) | 
{\cal H} | xz, \chi_- (\pi/2 + n \pi) \rangle 
= i (\sqrt{3}\lambda/2) (-1)^n$ 
and 
$\langle x^2-y^2,\chi_- (\pi/2 + n \pi) | 
{\cal H} | xz, \chi_- (\pi/2 + n \pi) \rangle = -i (\lambda/2) (-1)^n$]. 
%at $\theta = \pi/2 + n \pi$, 
%despite 
%in spite of 
%$|c_{x^2-y^2,-}^{x^2-y^2,-}(n\pi)| \sim 1$. 
%The relation  
%Here, 
In contrast, $|c_{x^2-y^2,-}^{x^2-y^2,-}(n\pi)| \sim 1$ 
is realized by 
%due to 
the small hybridization 
at $\theta = n \pi$ 
between the $d\gamma$ states 
through the $d\varepsilon$ states, 
%between the $d\gamma$ and $d\varepsilon$ states. 
%at $\theta = n \pi$. 
as found from the fact that 
%almost elements 
%in the matrix element of $\cal H}$ between 
most of 
$\langle j_1,\chi_- (n \pi) \rangle | {\cal H} |j_2,\chi_- (n \pi) \rangle$ 
are zero, 
where $j_1=x^2-y^2$ or $3z^2-r^2$, and $j_2 = xy$, $yz$, or $xz$.

\subsubsection{Crystal field of tetragonal symmetry}
\label{tetra_sym}

We consider 
the system with the crystal field of tetragonal symmetry. 
In Fig. \ref{c_coeff_D}, 
we show the $D_{\varepsilon,-}^{(d)}/D_{\gamma,-}^{(d)}$ 
($=r_{s,\sigma \to \varepsilon,-}/r_{s,\sigma \to \gamma,-}$) 
dependence of $C_2$ of Eq. (\ref{C2_EDM}) [$C_4$ of Eq. (\ref{C4_EDM})] 
for the system with 
$r_{s,+ \to \gamma 1,-}=r_{s,- \to \gamma 1,-}=0.01$ 
by the solid red [dashed red] curve. 
We find that 
the system 
shows 
$C_2 > 0$ and $C_4>0$ for $D_{\varepsilon,-}^{(d)}/D_{\gamma,-}^{(d)} < 1$, 
$C_2 = C_4=0$ at $D_{\varepsilon,-}^{(d)}/D_{\gamma,-}^{(d)}=1$, 
and $C_2 < 0$ and $C_4<0$ for $D_{\varepsilon,-}^{(d)}/D_{\gamma,-}^{(d)}>1$. 
By using 
Eq. (\ref{Tr-AMR0}), 
TAMR$(\theta)$ 
of Eq. (\ref{Tr-AMR(theta)}) 
is then rewritten as
\begin{eqnarray}
\label{Tr-AMR_tetra}
{\rm TAMR}(\theta) = 
C_2 (-1+  \cos 2 \theta)
+ 
C_4 (-1+  \cos 4 \theta). 
\end{eqnarray}
The system 
therefore exhibits 
the twofold and fourfold symmetric TAMR effect with 
${\rm TAMR}(\theta) \le 0$ 
for $D_{\varepsilon,-}^{(d)}/D_{\gamma,-}^{(d)} < 1$ 
and 
${\rm TAMR}(\theta) \ge 0$ 
for $D_{\varepsilon,-}^{(d)}/D_{\gamma,-}^{(d)} > 1$, 
while 
the system 
has 
${\rm TAMR}(\theta)= 0$ at $D_{\varepsilon,-}^{(d)}/D_{\gamma,-}^{(d)} = 1$.

\begin{figure}[ht]
\begin{center}
\includegraphics[width=.4\linewidth]{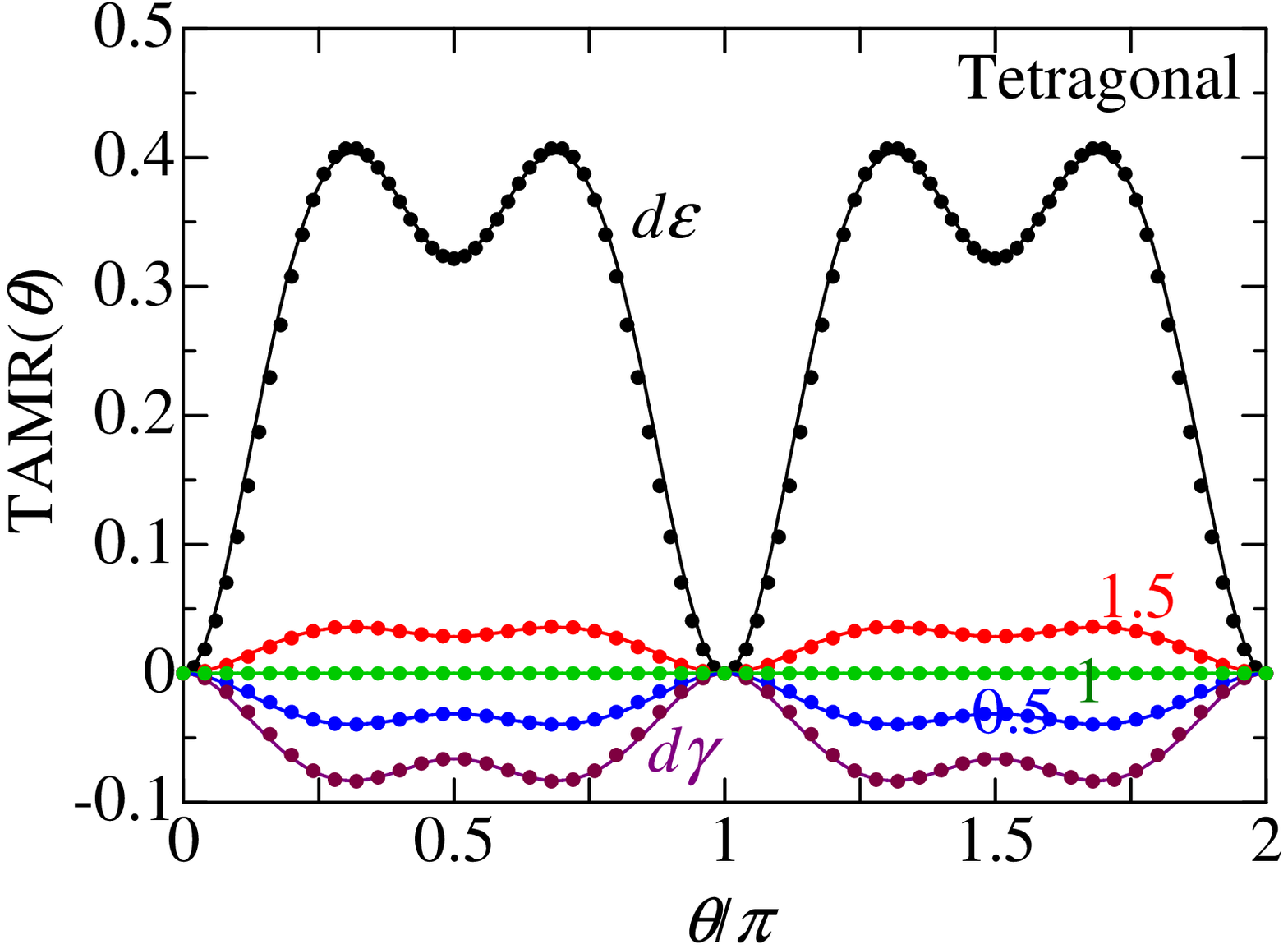} \\[-0.8cm]
\hspace*{-7cm}(a)\\
\vspace{0.3cm}
\includegraphics[width=.4\linewidth]{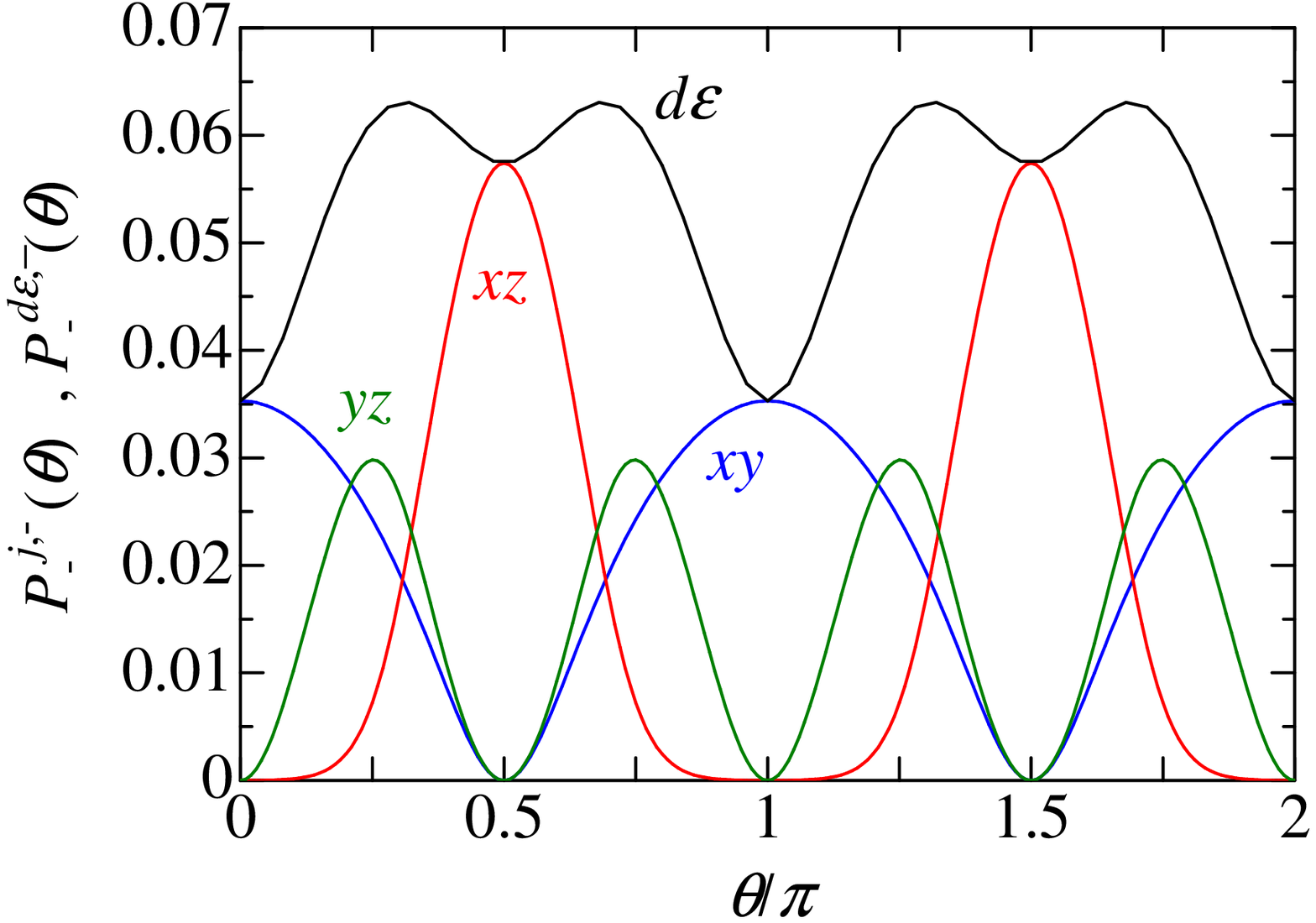} \\[-0.8cm]
\hspace*{-7cm}(b)\\
\vspace{0.3cm}
\includegraphics[width=.4\linewidth]{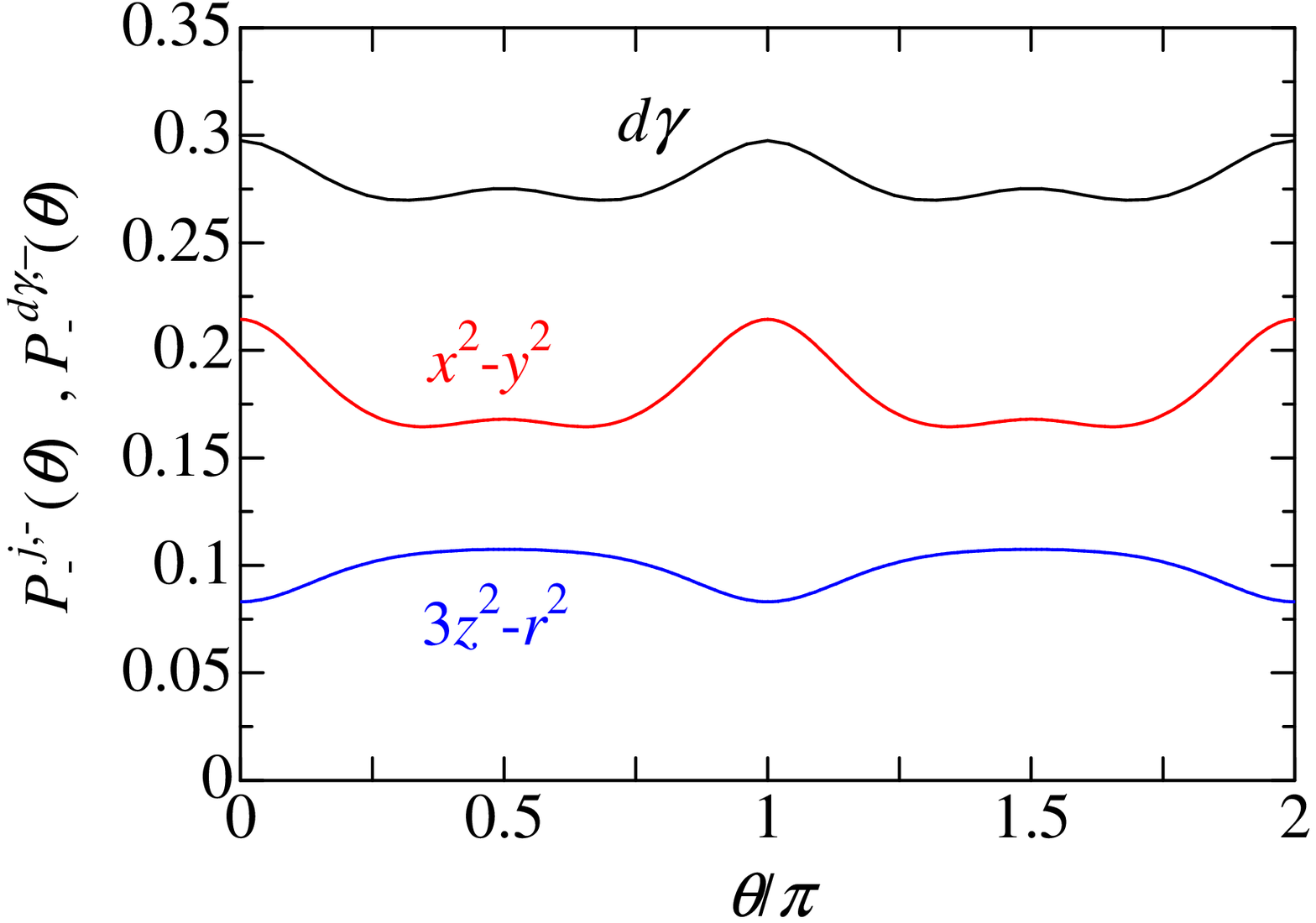} \\[-0.8cm]
\hspace*{-7cm}(c)\\
\vspace{0.3cm}
\includegraphics[width=.4\linewidth]{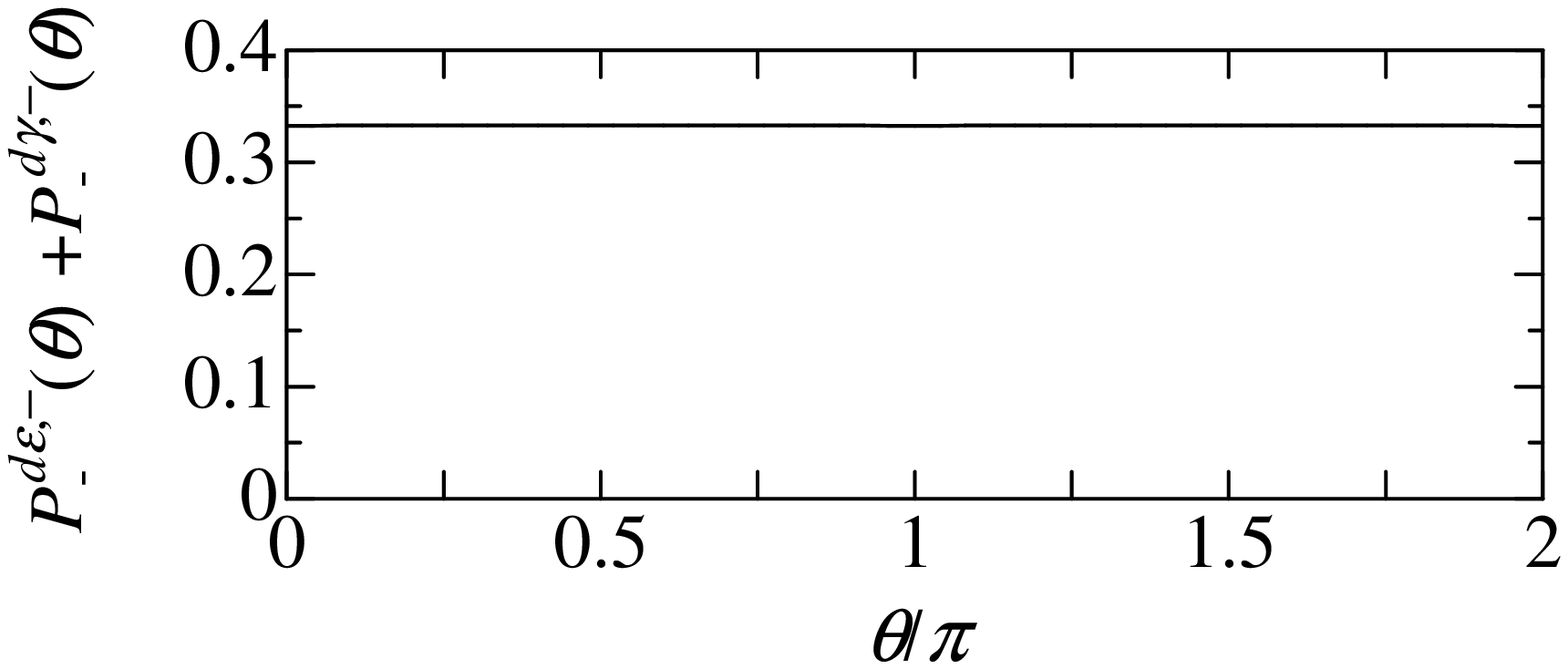} \\[-0.8cm]
\hspace*{-7cm}(d)\\
\vspace{0.3cm}
\caption{
(Color) 
The case of 
strong ferromagnets 
with a crystal field of tetragonal symmetry. 
The system has 
$D_{j,+}^{(d)} = 0$ (i.e., $r_{s,\sigma \to j,+}=0$), 
$r=0.001$, 
$H=1$ eV, 
$\lambda=\lambda'=0.05$ eV, 
$\Delta =0.1$ eV, 
and $\delta= 0.05$ eV. 
The notation is the same as in Fig. \ref{tr-amr_c}. 
In (a), however, 
the respective colored dots show 
TAMR$(\theta)$ 
of Eq. (\ref{Tr-AMR_tetra}) 
with 
the evaluated $C_2$s of Eq. (\ref{C2_EDM}) 
and $C_4$s of Eq. (\ref{C4_EDM}), 
where 
$C_2$ and $C_4$ 
for each model 
are given 
%summarized 
in Table \ref{tab_C2_C4}. 
Furthermore, 
$P_-^{d\varepsilon,-}(\theta) + P_-^{d\gamma,-}(\theta)$ in (d) 
is the sum of 
$P_-^{d\varepsilon,-}(\theta)$ in (b) 
and $P_-^{d\gamma,-}(\theta)$ in (c). 
}
\label{tr-amr_t}
\end{center}
\end{figure}

In Fig. \ref{tr-amr_t}(a), we show 
the $\theta$ dependence of TAMR($\theta$) of Eq. (\ref{TAMR}) 
for 
the $d\varepsilon$, $d\gamma$, and $d\varepsilon+d\gamma$ models. 
The curves represent 
results 
calculated directly by using 
the exact diagonalization method. 
The dots show 
the results obtained by substituting 
the evaluated $C_2$s of Eq. (\ref{C2_EDM}) 
and $C_4$s of Eq. (\ref{C4_EDM}) 
into 
TAMR$(\theta)$ 
of Eq. (\ref{Tr-AMR_tetra}), 
where 
$C_2$ and $C_4$ 
for each model 
are 
%summarized 
given in Table \ref{tab_C2_C4}. 
We find 
contradictory behavior of TAMR($\theta$) 
between 
the $d \varepsilon$ model 
and the $d \gamma$ model. 
Namely, the $d\varepsilon$ model 
exhibits ${\rm TAMR}(\theta)\ge 0$ 
and has two small peaks on a broad peak 
in one period (i.e., $0 \le \theta < \pi$). 
Specifically, 
%For details, 
the $d\varepsilon$ model takes 
the highest value 0.40 at 
$\theta \sim \pi/4 + n\pi/2$ (Ref. \citen{extreme}) 
and 
the lowest value 0 at 
$\theta = n\pi$, 
where $n=0, \pm1, \pm2, \cdots$. 
In contrast, 
the $d\gamma$ model 
shows ${\rm TAMR}(\theta)\le 0$ 
and 
has two small dips in a broad dip 
in one period (i.e., $0 \le \theta < \pi$). 
Specifically, 
%For details, 
the $d\gamma$ model has 
the highest value 0 
at $\theta = n \pi$ 
and 
the lowest value $-$0.083 
at $\theta \sim \pi/4 + n\pi/2$ (Ref. \citen{extreme}), 
where $n=0, \pm1, \pm2, \cdots$. 
The $d \varepsilon+d\gamma$ model 
takes TAMR($\theta$) between 
TAMR($\theta$) of the $d \varepsilon$ model and that of the $d\gamma$ model. 
Namely, the $d \varepsilon+d\gamma$ model has 
$0 \le {\rm TAMR(\theta)} \le 0.036 $ 
for $D_{\varepsilon,-}^{(d)}/D_{\gamma,-}^{(d)}=1.5$, 
${\rm TAMR(\theta)} = 0$ 
at $D_{\varepsilon,-}^{(d)}/D_{\gamma,-}^{(d)}=1$, 
and $-0.039 \le {\rm TAMR(\theta)} \le 0$ 
for $D_{\varepsilon,-}^{(d)}/D_{\gamma,-}^{(d)}=0.5$. 
In other words, 
the positive and large TAMR($\theta$) for the $d \varepsilon$ model 
is reduced 
by the negative TAMR($\theta$) for the $d\gamma$ model 
[see Eq. (\ref{TAMR_de+dg2})].

\begin{table}[ht]
\begin{center}
\caption{
The coefficients $C_2$ of Eq. (\ref{C2_EDM}) 
and $C_4$ of Eq. (\ref{C4_EDM}) 
for strong ferromagnets 
with a crystal field of tetragonal symmetry. 
The system has 
$D_{j,+}^{(d)} = 0$ (i.e., $r_{s,\sigma \to j,+}=0$), 
$r=0.001$, 
$H=1$ eV, 
$\lambda=\lambda'=0.05$ eV, 
$\Delta =0.1$ eV, 
and $\delta= 0.05$ eV. 
For these systems, we consider 
$d \varepsilon$, $d\gamma$, and $d\varepsilon+d \gamma$ models. 
In particular, 
the $d\varepsilon+d \gamma$ model has 
$D_{\varepsilon,-}^{(d)}/D_{\gamma,-}^{(d)}=0.5$, 1, and 1.5. 
Additionally, TAMR($\theta$) of Eq. (\ref{Tr-AMR_tetra}) 
with each set of $C_2$ and $C_4$ in this table 
is indicated by the dots 
in Fig. \ref{tr-amr_t}(a). 
}
\begin{tabular}{lcc}
\hline 
Model & $C_2$ & $C_4$ \\
\hline 
$d\varepsilon$ & $-$0.16 & $-$0.11\\
$d\gamma$ & 0.033 & 0.022  \\
$d\varepsilon+d \gamma$ ($D_{\varepsilon,-}^{(d)}/D_{\gamma,-}^{(d)}=0.5$) & 0.016 & 0.011 \\
$d\varepsilon+d \gamma$ ($D_{\varepsilon,-}^{(d)}/D_{\gamma,-}^{(d)}=1$) & 0 & 0 \\
$d\varepsilon+d \gamma$ ($D_{\varepsilon,-}^{(d)}/D_{\gamma,-}^{(d)}=1.5$) & $-$0.014 & $-$0.0096 \\
\hline
\end{tabular}
\label{tab_C2_C4}
\end{center}
\end{table}

We can explain 
the behavior of TAMR($\theta$) 
in the same way as for the cubic system. 
First, 
the $\theta$ dependences of 
TAMR($\theta$) for the $d\varepsilon$ and $d\gamma$ models 
directly reflect 
those of $P_-^{d\varepsilon,-}(\theta)$ 
and 
$P_-^{d\gamma,-}(\theta)$, 
respectively, 
as found from Eqs. (\ref{TAMR_approx})--(\ref{1/tau_approx}), 
(\ref{TAMR_depsilon}), and (\ref{TAMR_dgamma}). 
In Fig. \ref{tr-amr_t}(b) [(c)], 
we show the $\theta$ dependence of 
$P_-^{d\varepsilon,-}(\theta)$ [$P_-^{d\gamma,-}(\theta)$] 
by the black curve. 
In addition, Fig. \ref{tr-amr_t}(d) shows 
the $\theta$ dependence of 
$P_-^{d\varepsilon,-}(\theta)+ P_-^{d\gamma,-}(\theta)$. 
We find contradictory behavior between 
$P_-^{d\varepsilon,-}(\theta)$ and $P_-^{d\gamma,-}(\theta)$, 
in which 
$P_-^{d\gamma,-}(\theta)$ decreases (increases) 
as $P_-^{d\varepsilon,-}(\theta)$ increases (decreases). 
This behavior arises under 
the condition of 
$P_-^{d\varepsilon,-}(\theta)+ P_-^{d\gamma,-}(\theta) \approx 1/3$ 
of Eq. (\ref{1/3}). 
The condition can be 
confirmed 
in Fig. \ref{tr-amr_t}(d). 
Next, 
we look at 
%mention 
the value of TAMR($\theta$) 
for the $d\varepsilon + d\gamma$ model 
on the basis of 
the approximate expression of TAMR($\theta$) 
of Eqs. (\ref{TAMR_depsilon})--(\ref{TAMR_de+dg2}). 
%Eq. (\ref{TAMR_de+dg2}). 
% or (\ref{TAMR_de+dg1}). 
As seen from Eq. (\ref{relation_value}), 
%(\ref{de_mag}) [(\ref{dg_mag})],
%when $D_{\varepsilon,-}^{(d)}/D_{\gamma,-}^{(d)}=1.5$ [0.5], 
%the value of 
TAMR($\theta$) of Eq. (\ref{TAMR_de+dg2}) 
%[(\ref{TAMR_de+dg1})] 
is less than or equal to 
that for the $d\varepsilon$ model of Eq. (\ref{TAMR_depsilon}) 
and 
greater than or equal to 
%is  than or equal to 
that for the $d\gamma$ model of Eq. (\ref{TAMR_dgamma}). 
In particular, 
when $D_{\varepsilon,-}^{(d)}/D_{\gamma,-}^{(d)}=1$, 
TAMR($\theta$) of Eq. (\ref{TAMR_de+dg}) or (\ref{TAMR_de+dg1}) 
is 0 [see Eq. (\ref{0})].

We describe the details of $P_-^{d\varepsilon,-}(\theta)$ 
of Eq. (\ref{P_varepsilon}) 
for the $d\varepsilon$ model [see Fig. \ref{tr-amr_t}(b)]. 
The quantity 
$P_-^{d\varepsilon,-}(\theta)$ has 
peaks 
at $\theta \sim \pi/4 + n\pi/2$ 
like the cubic system, 
where $n=0, \pm1, \pm2, \cdots$. 
The peaks come from 
$P_-^{xy,-}(\theta)$, $P_-^{yz,-}(\theta)$, and $P_-^{xz,-}(\theta)$ 
at $\theta \sim \pi/4 + n\pi/2$. 
Namely, 
the $d \varepsilon$ states 
consisting of 
$|xy, \chi_- (\theta) \rangle$, 
$|yz, \chi_- (\theta) \rangle$, 
and 
$|xz, \chi_- (\theta) \rangle$ 
are strongly hybridized to the $d\gamma_{x,-}$ states 
at $\theta \sim \pi/4 + n\pi/2$. 
In addition, 
$P_-^{d\varepsilon,-}(n \pi)$ 
[$P_-^{d\varepsilon,-}(\pi/2 + n \pi)$] 
is formed by 
only $P_-^{xy,-}(n \pi)$ [$P_-^{xz,-}(\pi/2 + n \pi)$] 
like the cubic system, 
where $n=0, \pm1, \pm2, \cdots$. 
We note, however, that 
the tetragonal system shows 
$P_-^{xz,-}(\pi/2+ n \pi) > P_-^{xy,-}(n \pi)$, 
whereas the cubic system has 
$P_-^{xz,-}(\pi/2+ n \pi) = P_-^{xy,-}(n \pi)$ 
[see Fig. \ref{tr-amr_c}(b)]. 
The relation $P_-^{xz,-}(\pi/2+ n \pi) > P_-^{xy,-}(n \pi)$ indicates that 
$|xz, \chi_- (\theta) \rangle$ 
is strongly hybridized to the $d \gamma_{x,-}$ states 
compared with 
$|xy, \chi_- (\theta) \rangle$, 
because 
the energy differences between 
$|xz, \chi_- (\theta) \rangle$ 
and 
$d\gamma$ states 
are smaller than those between 
$|xy, \chi_- (\theta) \rangle$ 
and $d\gamma$ states, 
as shown in Fig. \ref{energy}. 
We also find 
$P_-^{xy,-}(n \pi) > P_-^{yz,-}(\pi/4 + n \pi/2)$ 
like the cubic system. 
This relation means that 
the probability density of the $d\gamma_{x,-}$ states in 
$|xy, \chi_- (n \pi) )$
is larger than 
that in $|yz, \chi_- (\pi/4 + n \pi/2) )$.

We give the details of $P_-^{d\gamma,-}(\theta)$ 
of Eq. (\ref{P_gamma}) 
for the $d\gamma$ model [see Fig. \ref{tr-amr_t}(c)]. 
The behavior of $P_-^{d\gamma,-}(\theta)$ 
mainly 
comes from 
that of $P_-^{x^2-y^2,-}(\theta)$ 
because of $P_-^{x^2-y^2,-}(\theta)>P_-^{3z^2 - r^2,-}(\theta)$ 
for the whole range of $\theta$. 
This feature is different from that for the cubic system. 
Namely, 
$P_-^{x^2-y^2,-}(\theta)>P_-^{3z^2 - r^2,-}(\theta)$ 
for the whole range of $\theta$ 
does not correspond to 
$P_-^{x^2-y^2,-}( n\pi)=P_-^{3z^2 - r^2,-}(\pi/2 + n\pi)$ 
for the cubic system. 
The relation 
$P_-^{x^2-y^2,-}(\theta)>P_-^{3z^2 - r^2,-}(\theta)$ 
shows that 
the probability density of 
$|x^2-y^2, \chi_- (\theta) \rangle$ 
of the $x$ direction 
is always larger than 
that of 
$|3z^2-r^2, \chi_- (\theta) \rangle$ 
of the $x$ direction. 
Roughly speaking, 
$P_{\sigma}^{j,\varsigma}(\theta)$ of Eq. (\ref{P_m}) 
is approximately 
expressed as 
$P_-^{x^2-y^2,-}(\theta) \approx  
(1/4) \left| c_{x^2-y^2,-}^{x^2-y^2,-}(\theta) \right|^2 \sim 1/4$ 
and 
$P_-^{3z^2-r^2,-}(\theta) \approx  
(1/12) \left| c_{3z^2-r^2,-}^{3z^2-r^2,-}(\theta) \right|^2 \sim 1/12$ 
under the assumption of 
$|c_{3z^2-r^2,-}^{x^2-y^2,-}(\theta)| \ll 
|c_{x^2-y^2,-}^{x^2-y^2,-}(\theta)| \sim 1$ 
and 
$|c_{x^2-y^2,-}^{3z^2-r^2,-}(\theta)| \ll 
|c_{3z^2-r^2,-}^{3z^2-r^2,-}(\theta)| \sim 1$, 
as found from the values of $P_-^{x^2-y^2,-}(\theta)$ 
and $P_-^{3z^2-r^2,-}(\theta)$ in Fig. \ref{tr-amr_t}(c). 
%[also see the values of $P_-^{x^2-y^2,-}(\theta)$ and $P_-^{3z^2-r^2,-}(\theta)$ in Fig. \ref{tr-amr_t}(c)]. 
This assumption may be valid 
%effective 
for the tetragonal system with small hybridization 
between nondegenerate $d\gamma$ states 
(also see the submatrix of the $d \gamma$ states 
in ${\cal H}$ of Table \ref{matrix}). 
We also find that 
$P_-^{d\gamma,-}(\theta)$ has peaks at $\theta = n \pi$, 
with $n=0, \pm1, \pm2, \cdots$. 
The peaks originate from those of $P_-^{x^2-y^2,-}(n \pi)$. 
Here, $|\psi_{x^2-y^2, -} (\theta) )$ 
includes the $d \gamma_{x,-}$ states 
with the highest probability density 
at $\theta =n\pi$. 
%In contrast, 
%that is, $|c_{x^2-y^2,-}^{x^2-y^2,-}(n\pi)| \sim 1$ is realized due to the small hybridization at $\theta = n \pi$ between the $d\gamma$ states through the $d\varepsilon$ states. 
%as found from the fact that most of $\langle j_1,\chi_- (n \pi) \rangle | {\cal H} |j_2,\chi_- (n \pi) \rangle$ are zero, where $j_1=x^2-y^2$ or $3z^2-r^2$, and $j_2 = xy$, $yz$, or $xz$.  

\section{Comment}
\label{comment}
On the basis of the above-mentioned results, 
we offer a comment on 
the experimental results of 
the temperature, $T$, dependences of $C_2$ and $C_4$ in 
TAMR$(\theta)$ 
of Eq. (\ref{TAMR}) 
for Fe$_4$N.\cite{Kabara1} 
Here, Fe$_4$N is considered to be 
a strong ferromagnet.\cite{SW_FM,Kokado1,Kokado3} 
In addition, the parameters in Sect. \ref{strong} 
are appropriate for Fe$_4$N.\cite{comment_param}

The noteworthy experimental result is 
the enhancement of $|C_2|$ and $|C_4|$ 
with a decrease in temperature for $T \lesssim 35$ K, 
i.e., 
(i) $C_2=0$ and $C_4=-0.005$ at $T \sim 35$ K 
and (ii) $C_2=-0.02$ and $C_4=-0.01$ at $T=5$ K.\cite{Kabara1} 
Result (i) 
agrees 
fairly well with 
$C_2=0$ and $C_4=-0.0048$ evaluated 
for the $d\varepsilon+d \gamma$ model 
of $D_{\varepsilon,-}^{(d)}/D_{\gamma,-}^{(d)}=1.5$ 
with a crystal field of ``cubic'' symmetry (see Table \ref{tab_C4}). 
In contrast, result (ii) 
corresponds well with 
$C_2=-0.014$ and $C_4=-0.0096$ evaluated 
for the $d\varepsilon+d \gamma$ model 
of $D_{\varepsilon,-}^{(d)}/D_{\gamma,-}^{(d)}=1.5$ 
with a crystal field of ``tetragonal'' symmetry (see Table \ref{tab_C2_C4}). 
%It is noted 
Note here that 
this $D_{\varepsilon,-}^{(d)}/D_{\gamma,-}^{(d)}$ (=1.5) 
is comparable to 
$1.1 \lesssim 
D_{\varepsilon,-}^{(d)}/D_{\gamma,-}^{(d)} 
\lesssim 1.4$ 
evaluated from our theoretical analysis of 
the experimental result of the in-plane AMR effect for 
$T \le 35$ K (see Fig. 11 of Erratum II in Ref. \citen{Kokado3}). 
From the above result, we predict that 
the tetragonal distortion may proceed 
with decreasing temperature for 
$T \lesssim 35$ K.

The tetragonal distortion has already been predicted 
also in our analysis of 
the experimental results of the in-plane AMR ratio 
for Fe$_4$N,\cite{Kokado3} 
where 
the experimental results show 
the 
appearance of $C_4$ 
for $T \le 35$ K.\cite{Tsunoda2} 
Concretely, 
by using the electron scattering theory 
based on the second-order perturbation theory, 
we showed that 
a model with a crystal field of tetragonal symmetry 
can reproduce the appearance of $C_4$.\cite{Kokado3}

On the other hand, 
Yahagi {\it et al.} 
investigated 
the fourfold symmetric AMR effect 
for a cubic single-crystal ferromagnetic model 
by using 
the Kubo formula 
and the multi-orbital d-impurity Anderson model, 
where 
the d states have the spin--orbit interaction and 
crystal field of cubic symmetry.\cite{Yahagi}  
They found that 
$C_4$ in the in-plane AMR ratio 
appears even for the cubic system 
as a result of the fourth-order perturbation 
with respect to the spin--orbit interaction. 
Here, 
the splitting of the d states 
due to the spin--orbit interaction 
is responsible for the appearance of $C_4$. 
Furthermore, 
by performing the numerical (i.e., nonperturbative) calculation for 
TAMR($\theta$) of the cubic system, 
they showed that $C_4$ appears 
%under 
by the same mechanism as 
the above-mentioned $C_4$ in the in-plane AMR ratio. 
Simultaneously, they confirmed the absence of $C_2$ in TAMR($\theta$). 
The absence of $C_2$ 
is 
certainly valid from the viewpoint of the symmetry of the cubic crystal. 
In addition, 
the absence of $C_2$ 
for the cubic system is confirmed also in our present study 
based on the exact diagonalization method (see Sect. \ref{cubic_sym}).

We emphasize, however, that 
the experimental result of TAMR($\theta$) for Fe$_4$N shows 
$C_2 \ne 0$ for $T \lesssim 35$ K.\cite{Kabara1} 
This $C_2 \ne 0$ is actually obtained 
for the tetragonal system 
with the finite $\delta$ of Eq. (\ref{delta}) 
in our present study 
(see Sect. \ref{tetra_sym}). 
Therefore, we guess that 
the tetragonal distortion may be necessary 
to explain 
the experimental results of 
both TAMR($\theta$) and the in-plane AMR ratio for Fe$_4$N.

\section{Conclusion}
\label{sec_conc}
We developed a theory of 
the anisotropic magnetoresistance effects 
of arbitrary directions of 
${\mbox{\boldmath $I$}}$ 
and 
${\mbox{\boldmath $M$}}$ 
for ferromagnets. 
Here, we used the electron scattering theory 
with $s$--$d$ scattering processes. 
The $s$--$d$ scattering means 
that the conduction electron is scattered into the localized d states 
by a nonmagnetic impurity. 
The resistivity due to the $s$--$d$ scattering 
directly reflects 
the probability density of the d states 
of the ${\mbox{\boldmath $I$}}$ direction. 
The d states are numerically obtained by 
applying the exact diagonalization method 
to the Hamiltonian of the d states 
with the exchange field, crystal field, and spin--orbit interaction. 
Using this theory, 
we investigated the TAMR effect 
for strong ferromagnets 
with a crystal field of cubic or tetragonal symmetry. 
The cubic system exhibits the fourfold TAMR effect, 
%while 
whereas 
the tetragonal system shows 
the twofold and fourfold TAMR effect. 
In addition, 
the cubic or tetragonal system has 
${\rm TAMR}(\theta) \le 0$ 
for $D_{\varepsilon,-}^{(d)}/D_{\gamma,-}^{(d)} < 1$, 
${\rm TAMR}(\theta)= 0$ at $D_{\varepsilon,-}^{(d)}/D_{\gamma,-}^{(d)} = 1$, 
and 
${\rm TAMR}(\theta) \ge 0$ 
for $D_{\varepsilon,-}^{(d)}/D_{\gamma,-}^{(d)} > 1$. 
We also found 
%the 
contradictory behavior of TAMR($\theta$) 
between 
the $d \varepsilon$ model with 
$D_{\varepsilon,-}^{(d)} \ne 0$ and $D_{\gamma,-}^{(d)}=0$ 
and 
the $d \gamma$ model with 
$D_{\varepsilon,-}^{(d)} = 0$ and $D_{\gamma,-}^{(d)}\ne0$. 
Such behavior appears under the condition for 
the probability density of the d states 
of the ${\mbox{\boldmath $I$}}$ direction. 
Finally, on the basis of 
the calculation results, 
we commented on the experimental results of 
the TAMR effect for Fe$_4$N, i.e., 
the enhancement of $|C_2|$ and $|C_4|$ 
with a decrease in temperature. 
We predicted that 
the tetragonal distortion may proceed 
with decreasing temperature.

\acknowledgments
We would like to thank Mr. Yuta Yahagi of Tohoku University 
for useful discussion. 
This work has been supported by 
the Cooperative Research Project (H31/A06) of 
the RIEC, Tohoku University, 
and 
Grants-in-Aid for Scientific Research (C) (No. 19K05249) 
and (B) (No. 20H02177) 
from the Japan Society for the Promotion of Science.

\appendix

\section{TAMR($\theta$) and $P_{\sigma}^{j,\varsigma}(\theta)$}
\label{relation}

We describe the relations between 
TAMR($\theta$) and $P_{\sigma}^{j,\varsigma}(\theta)$ 
for the $d \varepsilon$, $d\gamma$, and $d \varepsilon+d\gamma$ models 
of a strong ferromagnet 
with $D_{j,+}^{(d)} = 0$, $\sum_j D_{j,-}^{(d)} \ne 0$, 
$r \ll 1$, and $r_{s,\sigma \to j,-} \ll 1$, 
where 
the actual parameters are given in Sect. \ref{model_param}. 
To begin with, 
we simply write 
TAMR($\theta$) of Eq. (\ref{TAMR}) 
for this system 
as
\begin{eqnarray}
\label{TAMR_approx}
{\rm TAMR}(\theta) 
\approx \frac{\rho_- (\theta) - \rho_- (0)}
{\rho_- (0)}. 
\end{eqnarray}
Here, we have used 
\begin{eqnarray}
\label{rho_approx}
\rho (\theta) =  \frac{\rho_+ (\theta) \rho_- (\theta)}
{\rho_+ (\theta) + \rho_- (\theta)} 
= \frac{\rho_+ (\theta) \rho_- (\theta)}
{\rho_+ (\theta) [ 1 + \rho_- (\theta)/\rho_+ (\theta)]}
\approx \rho_- (\theta), 
\end{eqnarray}
which is obtained 
on the basis of $\rho_- (\theta)/\rho_+ (\theta) \ll 1$.\cite{comment_rho} 
The angle $\theta$ dependence of 
TAMR($\theta$) of Eq. (\ref{TAMR_approx}) thus reflects 
that of $\rho_- (\theta)$. 
The angle $\theta$ dependence of $\rho_- (\theta)$ 
results from 
that of 
$\sum_j \sum_\varsigma 1/\tau_{s,- \to j,\varsigma}(\theta)$ 
of Eqs. (\ref{s-d_general}) and (\ref{no_phi35}) 
[also see Eqs. (\ref{rho_sigma^i}) and (\ref{tau_inv})]. 
We now substitute $D_{j,+}^{(d)} =0$ 
into 
$\sum_j \sum_\varsigma 1/\tau_{s,- \to j,\varsigma}(\theta)$. 
In addition, 
we take into account 
$D_{\varepsilon,-}^{(d)} \ne 0$ 
and 
$D_{\gamma,-}^{(d)} = 0$ 
for the $d \varepsilon$ model, 
$D_{\varepsilon,-}^{(d)} = 0$ 
and $D_{\gamma,-}^{(d)}\ne 0$ 
for the $d \gamma$ model, 
and 
$D_{\varepsilon,-}^{(d)} \ne 0$ 
and $D_{\gamma,-}^{(d)} \ne 0$ 
for the $d \varepsilon + d \gamma$ model. 
As a result, we can express 
$\sum_j \sum_\varsigma 1/\tau_{s,- \to j,\varsigma}(\theta)$ 
of Eqs. (\ref{s-d_general}) and (\ref{no_phi35}) as
\begin{eqnarray}
&&\sum_j \sum_\varsigma 
\frac{1}{\tau_{s,- \to j,\varsigma}(\theta)} 
= \frac{2\pi}{\hbar} n_{\rm imp} N_{\rm n} v_-^2 
\sum_j P_-^{j,-}(\theta) D_{j,-}^{(d)} \nonumber \\
&&\hspace*{0.5cm}=
\left\{
\begin{array}{l}
\label{1/tau_approx}
\displaystyle{\frac{2\pi}{\hbar}} n_{\rm imp} N_{\rm n} v_-^2 
P_-^{d\varepsilon,-}(\theta) D_{\varepsilon,-}^{(d)},\hspace{0.8cm}
{\rm for}\hspace{0.3cm}d\varepsilon~{\rm model}, \\
\displaystyle{\frac{2\pi}{\hbar}} n_{\rm imp} N_{\rm n} v_-^2 
P_-^{d\gamma,-}(\theta) D_{\gamma,-}^{(d)},\hspace{0.8cm}
{\rm for}\hspace{0.3cm}d\gamma~{\rm model}, \\
\displaystyle{\frac{2\pi}{\hbar}} n_{\rm imp} N_{\rm n} v_-^2 
\left[ 
P_-^{d\varepsilon,-}(\theta) D_{\varepsilon,-}^{(d)}
+ P_-^{d\gamma,-}(\theta) D_{\gamma,-}^{(d)} \right],\hspace{0.8cm}
{\rm for}\hspace{0.3cm}d\varepsilon + d\gamma~{\rm model}, 
\end{array}
\right.
\end{eqnarray}
where 
$P_-^{d\varepsilon,-}(\theta)$ 
[$P_-^{d\gamma,-}(\theta)$] 
is given by 
%have been given by 
Eq. (\ref{P_varepsilon}) [(\ref{P_gamma})], 
and 
$D_{\varepsilon,-}^{(d)}$ [$D_{\gamma,-}^{(d)}$] 
is given by Eq. (\ref{DOS_e}) [(\ref{DOS_g})].

\section{TAMR($\theta$) for $d\varepsilon$, $d\gamma$, 
and $d\varepsilon+d\gamma$ models}
\label{TAMR_egeg}

Using Eqs. (\ref{TAMR_approx}), (\ref{1/tau_approx}), 
(\ref{r_s-s})--(\ref{1/tau_m_pm}), 
(\ref{rho_sigma^i}), and (\ref{tau_inv}), 
we 
%approximately 
obtain 
the approximate expression of TAMR($\theta$) 
for the $d\varepsilon$, $d\gamma$, and $d\varepsilon+d\gamma$ models 
of a strong ferromagnet. 
The expression of TAMR($\theta$) for the $d\varepsilon$ model 
is given by
\begin{eqnarray}
\label{TAMR_depsilon}
{\rm TAMR}(\theta) 
=\frac{3 \left[P_-^{d\varepsilon,-}(\theta) - P_-^{d\varepsilon,-}(0) 
\right] r_{s,- \to \varepsilon,-} }
{r + 3 P_-^{d\varepsilon,-}(0) r_{s,- \to \varepsilon,-} } 
\equiv {\rm TAMR}_{d\varepsilon}(\theta). 
\end{eqnarray}
The expression of TAMR($\theta$) for the $d\gamma$ model 
is 
\begin{eqnarray}
\label{TAMR_dgamma}
{\rm TAMR}(\theta) 
=\frac{3 \left[P_-^{d\gamma,-}(\theta) - P_-^{d\gamma,-}(0) 
\right] r_{s,- \to \gamma,-} }
{r + 3 P_-^{d\gamma,-}(0) r_{s,- \to \gamma,-} } 
\equiv {\rm TAMR}_{d\gamma}(\theta). 
\end{eqnarray}
The expression of TAMR($\theta$) for the $d\varepsilon + d\gamma$ model 
is 
\begin{eqnarray}
\label{TAMR_de+dg}
{\rm TAMR}(\theta) 
=\frac{3 \left[P_-^{d\varepsilon,-}(\theta) - P_-^{d\varepsilon,-}(0) 
\right] (r_{s,- \to \varepsilon,-} - r_{s,- \to \gamma,-}) }
{r + 3 P_-^{d\varepsilon,-}(0) r_{s,- \to \varepsilon,-} + \left[ 1 - 3 P_-^{d\varepsilon,-}(0)\right] r_{s,- \to \gamma,-}}
\equiv {\rm TAMR}_{d\varepsilon+d\gamma}(\theta), \nonumber \\
\end{eqnarray}
or
\begin{eqnarray}
\label{TAMR_de+dg1}
{\rm TAMR}(\theta) 
=\frac{3 \left[P_-^{d\gamma,-}(\theta) - P_-^{d\gamma,-}(0) 
\right] (r_{s,- \to \gamma,-} - r_{s,- \to \varepsilon,-}) }
{r + 3 P_-^{d\gamma,-}(0) r_{s,- \to \gamma,-} + \left[ 1 - 3 P_-^{d\gamma,-}(0)\right] r_{s,- \to \varepsilon,-}}
\equiv {\rm TAMR}_{d\varepsilon+d\gamma}(\theta), \nonumber \\
\end{eqnarray}
where 
the condition of Eq. (\ref{1/3}) has been used. 
%In addition, 
This ${\rm TAMR}(\theta)$ 
%${\rm TAMR}_{d\varepsilon+d\gamma}(\theta)$ 
of Eq. (\ref{TAMR_de+dg}) or (\ref{TAMR_de+dg1}) 
can be rewritten as
%is rewritten as
\begin{eqnarray}
\label{TAMR_de+dg2}
%{\rm TAMR}_{d\varepsilon+d\gamma}(\theta) 
{\rm TAMR}(\theta) 
=
(1-X){\rm TAMR}_{d\varepsilon}(\theta)
+ (1-Y){\rm TAMR}_{d\gamma}(\theta) 
\equiv {\rm TAMR}_{d\varepsilon+d\gamma}(\theta),
\end{eqnarray}
with
\begin{eqnarray}
X= \displaystyle{
\frac{
\left[1-3 P_-^{d\varepsilon,-}(0) \right] r_{s,- \to \gamma,-}}
{ r + 3P_-^{d\varepsilon,-}(0)r_{s,- \to \varepsilon,-}
+ \left[1-3 P_-^{d\varepsilon,-}(0) \right] r_{s,- \to \gamma,-}}}, \\
Y=\displaystyle{
\frac{
\left[1-3 P_-^{d\gamma,-}(0) \right] r_{s,- \to \varepsilon,-}}
{ r + 3P_-^{d\gamma,-}(0)r_{s,- \to \gamma,-}
+ \left[1-3 P_-^{d\gamma,-}(0) \right] r_{s,- \to \varepsilon,-}}},
\end{eqnarray}
where $0 \lesssim X < 1$ and $0 \lesssim Y <1$. 
The range of $X$ [$Y$] 
%In Eqs. (\ref{TAMR_de+dg}) and (\ref{TAMR_de+dg1}), 
%we have 
is given by using 
%obtained on the basis of 
%obtained from 
$1 - 3 P_-^{d\varepsilon,-}(0) \gtrsim 0$ 
[$1 - 3 P_-^{d\gamma,-}(0) \gtrsim 0$] 
%under
%based on 
obtained from the condition of Eq. (\ref{1/3}). 
Here, 
when $r_{s,- \to \gamma,-}=0$ 
[i.e., $D_{\gamma,-}^{(d)}=0$], 
%[$r_{s,- \to \varepsilon,-}=0$ (i.e., $D_{\varepsilon,-}^{(d)}=0$)], 
${\rm TAMR}_{d\varepsilon+d\gamma}(\theta)$ of Eq. (\ref{TAMR_de+dg2}) 
reduces to 
%becomes 
${\rm TAMR}_{d\varepsilon}(\theta)$ of Eq. (\ref{TAMR_depsilon}). 
%[${\rm TAMR}_{d\gamma}(\theta)$ of Eq. (\ref{TAMR_dgamma})]. 
When 
%$r_{s,- \to \gamma,-}=0$ (i.e., $D_{\gamma,-}^{(d)}=0$) 
$r_{s,- \to \varepsilon,-}=0$ [i.e., $D_{\varepsilon,-}^{(d)}=0$], 
${\rm TAMR}_{d\varepsilon+d\gamma}(\theta)$ of Eq. (\ref{TAMR_de+dg2}) 
%reduces to 
becomes 
%${\rm TAMR}_{d\varepsilon}(\theta)$ of Eq. (\ref{TAMR_depsilon}) 
${\rm TAMR}_{d\gamma}(\theta)$ of Eq. (\ref{TAMR_dgamma}). 
In addition, 
${\rm TAMR}_{d\varepsilon+d\gamma}(\theta)$ approaches 
${\rm TAMR}_{d\varepsilon}(\theta)$ 
%[${\rm TAMR}_{d\gamma}(\theta)$] 
with increasing 
$r_{s,- \to \varepsilon,-}/r_{s,- \to \gamma,-}$ 
[i.e., $D_{\varepsilon,-}^{(d)}/D_{\gamma,-}^{(d)}$], 
%$r_{s,- \to \varepsilon,-}$ [$\propto D_{\varepsilon,-}^{(d)}$], 
%[decreasing] 
%$D_{\varepsilon,-}^{(d)}/D_{\gamma,-}^{(d)}$, 
while 
${\rm TAMR}_{d\varepsilon+d\gamma}(\theta)$ approaches 
%${\rm TAMR}_{d\varepsilon}(\theta)$ 
${\rm TAMR}_{d\gamma}(\theta)$ 
with decreasing 
$r_{s,- \to \varepsilon,-}/r_{s,- \to \gamma,-}$. 
%[i.e., $D_{\varepsilon,-}^{(d)}/D_{\gamma,-}^{(d)}$]. 
%$r_{s,- \to \varepsilon,-}$ [$\propto D_{\varepsilon,-}^{(d)}$]. 
%$D_{\varepsilon,-}^{(d)}/D_{\gamma,-}^{(d)}$. 
In the case of 
${\rm TAMR}_{d\varepsilon}(\theta)\ge 0$ 
and 
${\rm TAMR}_{d\gamma}(\theta)\le 0$ 
%as 
shown in Figs. \ref{tr-amr_c}(a) and \ref{tr-amr_t}(a), 
we obtain the following relation 
from Eq. (\ref{TAMR_de+dg2}): 
\begin{eqnarray}
\label{relation_value}
{\rm TAMR}_{d\gamma}(\theta) 
\le {\rm TAMR}_{d\varepsilon+d\gamma}(\theta) 
\le {\rm TAMR}_{d\varepsilon}(\theta), 
\end{eqnarray}
regardless of 
$r_{s,- \to \varepsilon,-}/r_{s,- \to \gamma,-}$ 
[i.e., $D_{\varepsilon,-}^{(d)}/D_{\gamma,-}^{(d)}$]. 
%We note here that ${\rm TAMR}_{d\varepsilon+d\gamma}(\theta)$ approaches ${\rm TAMR}_{d\varepsilon}(\theta)$ [${\rm TAMR}_{d\gamma}(\theta)$] with increasing [decreasing] $D_{\varepsilon,-}^{(d)}/D_{\gamma,-}^{(d)}$. 
From Eq. (\ref{TAMR_de+dg}) or (\ref{TAMR_de+dg1}), 
we also have 
\begin{eqnarray}
\label{0}
{\rm TAMR}_{d\varepsilon+d\gamma}(\theta)=0,
\end{eqnarray}
for $r_{s,- \to \varepsilon,-}=r_{s,- \to \gamma,-}$ 
[i.e., $D_{\varepsilon,-}^{(d)}=D_{\gamma,-}^{(d)}$]. 
%from Eq. (\ref{TAMR_de+dg}) or (\ref{TAMR_de+dg1}). 
%Such results 
Equations (\ref{relation_value}) and (\ref{0}) 
are utilized 
in Sects. \ref{cubic_sym} and \ref{tetra_sym} 
[also see Figs. \ref{tr-amr_c}(a) and \ref{tr-amr_t}(a)].

\section{Condition for 
$P_-^{d\varepsilon,-}(\theta)+ P_-^{d\gamma,-}(\theta)$}
\label{cond1}

%\textcolor{red}
%{
On the basis of 
the condition of 
$\sum_j \sum_\varsigma P_\sigma^{j,\varsigma} (\theta)=1/3$ 
of Eq. (\ref{p_const1}), 
we derive the condition for 
$P_-^{d\varepsilon,-}(\theta)+ P_-^{d\gamma,-}(\theta)$ 
[see Eqs. (\ref{P_varepsilon}), (\ref{P_gamma}), and (\ref{P_m})]. 
We now focus on 
$c_{i,\sigma}^{j,\varsigma} (\theta)$ 
%the dominant components 
in $|\psi_{j,\varsigma} (\theta))$ 
%in $|\psi_{j,-} (\theta))$ 
of Eqs. (\ref{|m,chi_s)}) and (\ref{no_phi1}) 
under the present parameters with 
$\lambda/H \ll \lambda/\Delta < 1$ for the cubic system 
or 
$\lambda/H \ll \lambda/\Delta < \lambda/ \delta=1$ for the tetragonal system, 
where the actual parameters are given in Sect. \ref{model_param}. 
%Namely, $|\psi_{j,-} (\theta))$ of Eqs. (\ref{|m,chi_s)}), (\ref{no_phi1}), (\ref{no_phi3}), and (\ref{no_phi2}) is approximately expressed by using only the down spin components of $\sigma=-$: 
%\begin{eqnarray}
%\label{|m,-)}
%&&|\psi_{j,-} (\theta)) \approx 
%\sum_{\substack{i=xy, yz, xz\\x^2-y^2, 3z^2-r^2}} 
%c_{i,-}^{j,-} (\theta) 
%|i,\chi_- (\theta) \rangle. 
%\end{eqnarray}
Here, we have the relation 
$|c_{i,-}^{j,+} (\theta)| \ll |c_{i,-}^{j,-} (\theta)|$ 
%We note here that 
%$|c_{i,+}^{j,-} (\theta)|$ is much smaller than 
%$|c_{i,-}^{j,-} (\theta)|$ 
because 
%the 
mixing between the different spin states 
%arises 
occurs
between the states 
separated by the energy differences with $H$, 
while 
%the 
mixing between the same spin states 
occurs 
%arises 
between the states 
separated by the energy differences with 
$\Delta$ 
and/or 
$\delta$ 
(see ${\cal H}$ of Table \ref{matrix}). 
%Using 
From this relation and Eq. (\ref{P_m}), 
%Using Eqs. (\ref{|m,-)}) and (\ref{P_m}), 
we may then write 
the left-hand side of Eq. (\ref{p_const1}), 
$
\sum_{j}
\sum_\varsigma P_-^{j,\varsigma} (\theta)$, as
%by 
\begin{eqnarray}
\label{222}
&&
\sum_{\substack{j=xy, yz, xz\\x^2-y^2, 3z^2-r^2}} \sum_{\varsigma=+, -}
P_-^{j,\varsigma} (\theta)
\approx 
\sum_{\substack{j=xy, yz, xz\\x^2-y^2, 3z^2-r^2}}
P_-^{j,-} (\theta).
\end{eqnarray}
%From 
Using Eqs. (\ref{p_const1}) and (\ref{222}), 
we approximately obtain 
\begin{eqnarray}
\label{1/3}
&&\sum_{\substack{j=xy, yz, xz\\x^2-y^2, 3z^2-r^2}}
P_-^{j,-} (\theta) 
=
P_-^{d\varepsilon,-}(\theta)+ P_-^{d\gamma,-}(\theta)
\approx \frac{1}{3}. 
\end{eqnarray}
Equation (\ref{1/3}) is 
the condition for 
$P_-^{d\varepsilon,-}(\theta)+ P_-^{d\gamma,-}(\theta)$. 
%}


\begin{thebibliography}{9}
\bibitem{comment1}
%\textcolor{red}
%{
%As to 
In the 
%transformation 
calculation process 
from 
Eq. (\ref{tau_sd_inv}) to 
%$P_{\sigma}^{j,\varsigma}(\Theta,\Phi;\theta,\phi)$ in 
Eq. (\ref{s-d_general}), 
we perform the integration 
with respect to ${\mbox{\boldmath $r$}}$ 
in $(\psi_{j,\varsigma} (\theta,\phi)|e^{i{\mbox{\boldmath $k$}}_\sigma \cdot {\mbox{\boldmath $r$}}},\chi_\sigma (\theta,\phi) \rangle$. 
%Performing the integration with respect to ${\mbox{\boldmath $r$}}$ in $(\psi_{j,\varsigma} (\theta,\phi)|e^{i{\mbox{\boldmath $k$}}_\sigma \cdot {\mbox{\boldmath $r$}}},\chi_\sigma (\theta,\phi) \rangle$, 
%$(\psi_{j,\varsigma} (\theta,\phi)|e^{i{\mbox{\boldmath $k$}}_\sigma \cdot {\mbox{\boldmath $r$}}},\chi_\sigma (\theta,\phi) \rangle$ in Eq. (\ref{tau_sd_inv}) 
%can be calculated by using 
The result of the integration was given by Eq. (C.1) in Ref. \citen{Kokado3}.
\bibitem{comment_sigma}
%\textcolor{red}
%{
In Sects. 2.5.1 and 3.1 in Ref. \citen{Kokado1} 
and Sect. 3.1 in Ref. \citen{Kokado3}, 
we used a simple model with $n_+=n_-$, $m_+^* = m_-^*$, and $v_+ = v_-$, 
where 
$v_+^2$ ($v_-^2$) corresponds to 
%$v_+/\sqrt{3}$ ($v_-/\sqrt{3}$) corresponds to 
$3|V_{s\uparrow \to d\uparrow}|^2$ 
($3|V_{s\downarrow \to d\downarrow}|^2$) in Eq. (21) in Ref. \citen{Kokado1}. 
In Ref. \citen{Berger}, the author set $n_+ = n_-$ and $m_+^* = m_-^*$, 
%were set, 
and 
ignored the $\sigma$ dependence of parts corresponding to $v_\sigma$.
\bibitem{comment_param}
%\textcolor{red}
%{
These parameters are close to 
those used in a theoretical analysis for Fe$_4$N in Ref. \citen{Kokado3}. 
The parameters for Fe$_4$N are 
$H=1$ eV, 
$\lambda = \lambda'= 0.052$ eV, 
$\Delta = 0.1$ eV, 
%and 
$r = 0.0016$, 
and $r_{s,\sigma \to j,-}=0.01$, 
%and $r_{s,\sigma \to j,\varsigma}=0.01$, 
where 
$r_{s,\sigma \to j,-}$ of Eq. (\ref{r_s-d}) 
corresponds to 
$r_{i,-}$ of Eq. (68) 
%$\rho_{s \ro d \downarrow}/\rho{s\uparrow}$ in Sect. 3.3 
in Ref. \citen{Kokado3}. 
%\cite{Kokado3} 
%In particular, 
Here, 
$r$ was evaluated from 
%are the respective theoretical values evaluated from 
analyses using a combination of 
the first principles calculation and the Kubo formula 
(see Table I in Ref. \citen{Kokado1}). 
In addition, 
$r_{s,\sigma \to j,-}$ was evaluated to be 
$0.01 \lesssim r_{s,\sigma \to j,-} \lesssim 0.5$ 
%by comparing 
%from a theoretical analysis of 
from theoretical analyses of 
%from the theoretical analysis of 
the experimental result of the in-plane AMR ratio 
(see Sect. 3.3 in Ref. \citen{Kokado1}), 
where 
$r_{s,\sigma \to j,-}$ of Eq. (\ref{r_s-d}) 
corresponds to 
$\rho_{s \to d \downarrow}/\rho_{s\uparrow}$ in Sect. 3.3 
in Ref. \citen{Kokado1}.
\bibitem{comment_sd}
In Ref. \citen{Kokado1}, 
we obtained 
$\rho_{\perp, \sigma}$ of Eqs. (24) and (25) 
applying the perturbation theory to the ferromagnet with no crystal field. 
Here, 
$\rho_{\perp, \sigma}$ 
is 
a resistivity of the $\sigma$ spin 
for 
${\mbox{\boldmath $I$}} \perp {\mbox{\boldmath $M$}}$. 
From the expression of $\rho_{\perp, \sigma}$, 
we find that $\rho_{\perp, \sigma}$ is constant 
independent of the direction of ${\mbox{\boldmath $M$}}$, 
where ${\mbox{\boldmath $M$}}$ 
is rotatable 
in a plane perpendicular to ${\mbox{\boldmath $I$}}$. 
The system therefore 
has ${\rm TAMR(\theta)}=0$.
\bibitem{3d}
The d orbital states are given by 
$|xy \rangle = \frac{1}{\sqrt{2}i}(u_2 - u_{-2}) = xy f(r)$, 
$|yz \rangle = -\frac{1}{\sqrt{2}i}(u_1 + u_{-1}) = yz f(r)$, 
$|xz \rangle = -\frac{1}{\sqrt{2}}(u_1 - u_{-1}) = xz f(r)$, 
$|x^2-y^2 \rangle = 
\frac{1}{\sqrt{2}}(u_2 + u_{-2})=\frac{1}{2} (x^2-y^2) f(r)$, 
and 
$|3z^2-r^2 \rangle = u_0 = \frac{1}{2\sqrt{3}} (3z^2-r^2) f(r) $. 
Here, 
$u_m$ is the 3d orbital state with the magnetic quantum number $m$, 
where $m=-2$, $-1$, 0, 1, and 2.
\bibitem{extreme}
We can actually obtain the exact values of $\theta$ 
%which 
that take the extreme values 
of TAMR($\theta$) in Fig. \ref{tr-amr_t}. 
Here, TAMR($\theta$) is expressed by Eq. (\ref{Tr-AMR_tetra}), 
where the values of $C_2$ and $C_4$ are given in Table \ref{tab_C2_C4}. 
The values of $\theta$ are obtained 
from $\displaystyle{ \frac{d}{d\theta} {\rm TAMR}(\theta)=}0$. 
From $\displaystyle{ \frac{d}{d\theta} {\rm TAMR}(\theta)}=
-2 \sin 2\theta (C_2 + 4 C_4 \cos 2\theta ) =0$, 
we have 
$\theta = \displaystyle{\frac{1}{2}}\sin^{-1} 0$ 
or 
$\theta = 
\displaystyle{\frac{1}{2}}\cos^{-1} 
\left( - \displaystyle{\frac{C_2}{4C_4}} \right)$. 
In the $d \varepsilon$ model with $C_2 = -0.16$ and $C_4 = -0.11$, 
the values of $\theta$ 
are 
%$\theta = 
0, $0.31\pi$, $\pi/2$, and $0.69\pi$ 
for one period of $0 \le \theta < \pi$. 
The other models in Table \ref{tab_C2_C4} also have 
the same $\theta$ values as 
%the same values of $\theta$ as 
those above.
\bibitem{comment_rho}
On the basis of our previous study,\cite{Kokado3,Kokado5,Kokado6} 
we confirm $\rho_- (\theta)/\rho_+ (\theta) \ll 1$ 
for the strong ferromagnet 
with 
$D_{j,+}^{(d)} = 0$, $\sum_j D_{j,-}^{(d)} \ne 0$, 
$r \ll 1$, and $r_{s,\sigma \to j,-} \ll 1$ 
(see Sect. \ref{model_param}). 
Here, we have 
$\lambda/H \ll \lambda/\Delta < 1$ for the cubic system 
and 
$\lambda/H \ll \lambda/\Delta < \lambda/ \delta=1$ for the tetragonal system. 
We previously 
obtained the analytic expression of $\rho_\pm (\theta)$ 
for the in-plane AMR effect 
within 
the second-order perturbation theory.\cite{Kokado3,Kokado5,Kokado6} 
On the basis of the previous study [e.g., Eq. (3) in Ref. \citen{Kokado6}], 
we infer that 
an expression of $\rho_\pm (\theta)$ for the TAMR effect 
may be roughly expressed as 
$\rho_\pm (\theta)= 
\rho_{s,\pm} + \rho_{s,\pm \to d,\pm} 
+ [f_{1,\pm} + f_{2,\pm}(\theta)] \rho_{s,\pm \to d,\pm} 
+ [g_{1,\pm} + g_{2,\pm}(\theta)] \rho_{s,\pm \to d,\mp}$. 
Here, 
$\rho_{s,\sigma \to d,-}$ corresponds to $\rho_{s,\sigma \to j,-}$ 
of Eq. (\ref{rho_m_pm}), 
and 
$f_{1,\pm}+f_{2,\pm}(\theta)$ 
[$g_{1,\pm}+g_{2,\pm}(\theta)$] represents the coefficient of 
the $\rho_{s,\pm \to d,\pm}$ [$\rho_{s,\pm \to d,\mp}$] term, 
where 
$f_{1,\pm}$ and $g_{1,\pm}$ are 
the terms independent of $\theta$, 
%the constants independent of $\theta$ 
and 
$f_{2,\pm}(\theta)$ and $g_{2,\pm}(\theta)$ 
are the functions of $\theta$. 
In addition, 
$|f_{1,\pm}+f_{2,\pm}(\theta)|$ and $|g_{1,\pm}+g_{2,\pm}(\theta)|$ 
are considered to be at most 1 
when 
%under 
$\lambda/H \ll \lambda/\Delta < 1$ for the cubic system 
or 
$\lambda/H \ll \lambda/\Delta < \lambda/ \delta=1$ for the tetragonal system 
(also see ${\cal H}$ of Table \ref{matrix}). 
In the case of $D_{j,+}^{(d)} = 0$ (i.e., $\rho_{s,\sigma \to j,+}=0$), 
we have 
$\rho_+ (\theta)= 
\rho_{s,+} 
+ [g_{1,+} + g_{2,+}(\theta)] \rho_{s,+ \to d,-}$ 
and 
$\rho_- (\theta)= 
\rho_{s,-} + \rho_{s,- \to d,-} 
+ [f_{1,-} + f_{2,-}(\theta)] \rho_{s,- \to d,-}$. 
As a result, 
when $r \ll 1$ and $r_{s, \sigma \to j, -} \ll 1$ 
(i.e., $r_{s, \sigma \to d, -} \ll 1$), 
we obtain 
$\rho_-(\theta) /\rho_+(\theta) 
=
\left\{ r+r_{s,- \to d,-} + [f_{1,-} + f_{2,-} (\theta)] r_{s,- \to d,-}\right\} 
/\left\{ 1 + [g_{1,+} + g_{2,+}(\theta)] r_{s,+ \to d,-} \right\} 
\ll 1$, 
where 
$r_{s, \sigma \to d, -} = \rho_{s, \sigma \to d, -} / \rho_{s,+}$.
\end{thebibliography}
\end{document}